# Title: Simulation and assimilation of the digital human brain

**Authors:** Wenlian Lu [1,2,3,†], Xin Du [1,2,4,5,†], Jiexiang Wang [1,2,†], Longbin Zeng [1,2,†], Leijun Ye[1,2,†], Shitong Xiang [1,2,†], Qibao Zheng [1,2,†], Jie Zhang[1,2], Ningsheng Xu [1,6], Jianfeng Feng [1,2,3,4,7,8*] on behalf of the DTB Consortium

**Affiliations:**

[1] Institute of Science and Technology for Brain-Inspired Intelligence, Fudan University, Shanghai, China [2] Key Laboratory of Computational Neuroscience and Brain-Inspired Intelligence (Fudan University), Ministry of Education, China

[3] Shanghai Centre for Mathematical Sciences, Fudan University, Shanghai, China

[4] School of Computer Science, Fudan University, Shanghai, China

[5] School of Software Technology, Zhejiang University, Hangzhou, China

[6] State Key Laboratory of Integrated Chips and Systems, Fudan University, Shanghai, China

[7] Department of Computer Science, University of Warwick, Coventry, United Kingdom

[8] Zhangjiang Fudan International Innovation Centre, Shanghai, China

† These authors contributed equally to this work.

*Corresponding author. Email: jianfeng64@gmail.com

**Abstract**

Here, we present the Digital Brain (DB), a platform for simulating spiking neuronal networks at the large neuron scale of the human brain based on personalized magnetic-resonance-imaging data and biological constraints. The DB aims to reproduce both the resting state and certain aspects of the action of the human brain. An architecture with up to 86 billion neurons and 14,012 GPUs, including a two-level routing scheme between GPUs to accelerate spike transmission up to 47.8 trillion neuronal synapses, was implemented as part of the simulations. We show that the DB can reproduce blood-oxygen-level-dependent signals of the resting-state of the human brain with a high correlation coefficient, as well as interact with its perceptual input, as demonstrated in a visual task. These results indicate the feasibility of implementing a digital representation of the human brain, which can open the door to a broad range of potential applications.

## Introduction

Despite great progress in brain science, computer science and mathematics, simulating a spiking neuronal network of the human brain at a scale of up to 86 billion neurons remains a substantial challenge. Realistic simulations involve an intensively interconnected evolved structure[1,2] with long-range connections. A direct 'brute force' approach to simulating such a model on currently available high-performance computing systems is prone to fail due to its complexity and demand on resources[2,3,4]. Furthermore, direct measurement of the activity of billions of neurons in a human brain is not feasible either. Indeed, even with access to complete data on neuronal activity and resources to simulate a model on the scale of the human brain, there is the absence of mathematical tools that are needed to reverse engineer the complex neuronal network - a task that would involve fitting trillions of parameters. Thus, to produce a model to emulate the human brain as much as possible, and to contribute to studies of human intelligence, these challenges must be properly addressed.

In this paper, we demonstrate a methodology for constructing a model of the human brain at the full-scale of neuron numbers, and describe an explicit process that can infer the hyperparameters governing the entire system.

In previous research, the European Human Brain Project (HBP) aimed to create a digital infrastructure for neuroscience. For example, the SpiNNakker[5], as well as other simulators such as NEST [6], can perform simulations at the neuronal level of brains[7]. These approaches have been useful in, for example, simulating networks in both local circuits[8] and large-scale networks of multiple brain regions[9,10]. The Virtual Brain (TVB) also encompasses a large number of software tools, brain atlases, experimental datasets computational models and so forth, and is currently undergoing a large clinical trial on brain diseases[11,12]. However, there are still computational challenges to establish a digital brain: (1) how to efficiently simulate spiking neuronal networks at the large neuron scale of the human brain; (2) how to statistically infer such a "big" model by limited experimental data.

In the present work, we have implemented a Digital Brain (DB) platform to simulate the activity of the whole human brain, going beyond the mean field, by implementing a spiking neuronal network with up to 86 billion neurons and 47.8 trillion synapses. We divided the model into four main parts: cortex, subcortex, brainstem, and cerebellum, each having a different micro-architectural structure[1] (Figure 1a). A total of 49.52 trillion parameters were evaluated over 14,012 graphical processing unit (GPU) cards, each running at 1.10 GHz and having 16 GB Memory. This is a measure of the scale of computations required for a DB with 86 billion neurons, which also required the development of optimization algorithms for distributing neurons and managing spike communications within GPUs[13,14]. For simulations maintaining average neuronal firing rates of approximately 7Hz, 15Hz, and 30Hz, we have achieved a performance in which 1s of biological time requires 65s, 78.8s, and 118.8s of computation, respectively (Figure 1e), which corresponds to real-time factors of 65, 78.8 and 118.8, respectively. Combined with a hierarchical mesoscale data assimilation (HMDA) algorithm, the DB model was fitted to the functional-magnetic-resonance-imaging based blood-oxygen-level-dependent (fMRI-BOLD) signals of a human brain in resting state and in action, which indicates that the DB can serve as a platform for conducting various 'dry' digital experiments. As an example, a visual evaluation task was successfully performed using the DB (Figure 2f-i).

## Results

### *The architecture of the Digital Brain*

The challenges of the DB models include integrating multiple scales – meaning, incorporating microscopic features into mesoscopic and macroscopic models[15] – and accounting for the heterogeneity and biological details in brain architecture. These factors consistently increase the difficulty and complexity of computation, including the limitation of memory access and communication bandwidths when simulating a "big" model on general computing systems. Given our current computing capabilities, we strive to balance biological precision with affordable computing power. A full description of the architecture and operation of the DB is provided in the Supplementary Information (SI), and here we give an overview.

For this work, we scanned multimodal MRI data of the corresponding author of this paper (see Supplementary Section 1) and transformed them to a neuronal network architecture via the following characteristics (Figure 1a).

First, the number of neurons in each voxel was taken to be proportional to the gray matter volume obtained from T1-weighted MRI data. Specifically, they were distributed across different brain regions as follows: cortex (19,570 voxels), subcortex (2,191 voxels), brainstem (480 voxels), and cerebellum (1,622 voxels). Equivalently, our model assigned approximately 16.33 billion, , 0.64 billion , 0.05 billion , and 68.98 billion neurons to the cortex, subcortex, brainstem and cerebellum respectively.

Second, the structural connection probability between each pair of voxels was estimated by the row-normalized voxel-wise DWI (diffusion-weighted neuroimaging) matrix, meaning, each DWI element was driven by the sum of all elements at the same row, which characterizes the microstructure of white matter and the biophysical properties inferred from local diffusion properties. The proportion of voxel pairs with nonzero elements is quite low, at around 0.72% of all non-diagonal DWI elements. We assumed that long-range connections between neuronal populations are exclusively excitatory, as inhibitory connections typically exhibit more localized behaviour. Average synaptic inputs were set at 1000, 1000, 100, and 100 per neuron for the cortex, subcortex, brainstem, and cerebellum respectively.

Within every voxel in the cortex, distinct populations of excitatory (exc.) and inhibitory (inh.) neurons are present across the cortical layers L2/3, L4, L5, and L6, modeled as a micro-column[16]. Other voxels located in the subcortical region (subcortex, brainstem, and cerebellum) are represented by a randomly connected subnetwork involving two populations (exc. and inh.). Additionally, by combining positron emission tomography (PET) imaging of the synaptic vesicle glycoprotein 2A[17,18] with available neuroanatomy data, mean ratios of inner-voxel over all connections were set at around 2/7, 6/25, 14/25, and 16/125 for the cortex, subcortex, brainstem, and cerebellum respectively. Drawing upon this measured biological data (PET-DWI-sMRI) and combined with existing knowledge, we derived a weighted directed graph of neuronal populations with high heterogeneity, sparseness, and long-range cohesion (see Supplementary Section 2).

In the DB, each neuron was modelled using the Leaky Integrate and Fire (LIF) model, equipped with four types of synapses: ( α -amino-3-hydroxy-5-methyl-4-isoxazolepropionic-acid (AMAP), N-Methyl-D-aspartic-acid (NMDA), γ-Aminobutyric-acid-A ($GABA_A$), and $GABA_B$. These synapses exhibited an instantaneous jump followed by an exponential decay in their response properties. Each neuron received postsynaptic currents as input, along with independent Ornstein-Uhlenbeck (OU) processes - representing non-synaptic environmental noise. These inputs influenced the generation of action potentials (spikes) by the neuron model, the output describing the action potentials over time (see Supplementary Section 2). Hence, using the neuronal activities obtained from the simulation of the DB, we formulated the BOLD signal for each voxel using the Balloon-Windkessel model [19] (Figure 2a). To achieve a fit between empirical and simulated BOLD signals, the maximum synaptic conductance parameters of each neuron were estimated within the framework of HMDA[20] (Figure 2b). However, given the vast number of parameters involved, we assumed that the maximum synaptic conductance parameters for neurons of the same type within the same voxel followed the same distribution, thereby sharing equivalent hyperparameters that needed to be estimated. Notably, we demonstrated that highly consistent BOLD signals could be achieved by simulating a neuronal network of an arbitrarily larger scale than the assimilated model (Figs. 2a&b), where we set the neuron number to 200 million, and the average synaptic connection number per neuron to 100. When simulating a model with 86 billion neurons, the neuronal network was sampled from the same structure as the assimilated model, while the maximum synaptic conductance parameters were sampled from a distribution based on the estimated hyperparameters (see Methods and Supplementary Section 6).

*Simulation of the Digital Brain*

We performed a spiking neuronal network simulation of the human-brain scale, using 3503 computer nodes and 14012 GPUs. The whole brain model consists of 86 billion neurons and 47.8 trillion synapses with the topological structure described above (see Extended Data Table 1). In this model, the number of neurons loaded on each GPU was dominated by GPU memory size. Because the simulator stored enough neuron properties as configurable parameters to meet custom patterns, the data structure determined that up to 61.3 million neurons were loaded on a single GPU.

The DB platform is configured for client-server deployment. The server, housed within a high-performance computing (HPC) environment, serves as the back-end system supporting network simulation using our customized software. The storage system manages model files, including neuronal network representations, recorded data, and BOLD signals. Users can customize the network simulation parameters through a code interface (Figure 1b).

Due to memory access for handling extremely large computations and the communication of a substantial volume of spike packets occurring within milliseconds, the simulation cost of the DB was prohibitively expensive. Furthermore, we emphasize that heterogeneity of the brain's biological structure poses challenges for high-performance simulation on general high-performance computing (HPC) platforms. This heterogeneity, characterized by an uneven distribution of neurons and synapses, exacerbates issues related to memory access and communication imbalance, which are crucial factors posing challenges for high-performance simulation on general high-performance computing (HPC). To achieve low-latency communication between GPUs during DB simulation, we introduced a partitioning algorithm for determining the location of neurons on GPUs and a two-level routing method to optimize the data traffic and minimize the inter-GPU connections (Figure 1c and Supplementary Section 4). Compared with the natural partitioning method, where neurons are allocated to GPU cards one by one until reaching the physical memory limit, we achieved a substantial reduction of up to approximately 50% in the maximum inter-GPU traffic value within the system[13,14]. Using these methods, we simulated a whole human brain, comprising up to 86 billion neurons and 47.8 trillion synapses and achieved real-time factors of 65, 78.8, and 118.8 in simulation for average firing rates of around 7 Hz, 15 Hz and 30 Hz, respectively. This performance surpasses existing works, as demonstrated in Figure 1e and Supplementary Table 2. More details on the strong and weak scaling experiments, can be found in Supplementary Figures. 4-7.

We have validated the precision of DB simulation by two aspects. First, we inserted an independent small neuronal graph (referred to as a debug graph) into the whole brain network and compared the simulation results of the debug graph with results obtained from the same network deployed on a CPU system with the same computation and communication logic. We have found that the spike emission times computed from both methods were perfectly consistent in timing (measured in milliseconds), with relative absolute errors of membrane potentials less than $10^{-4}$ over a one-second simulation period. Second, the relative errors of membrane potentials of the Euler-Maruyama method using a time step of 1 millisecond over the method of a time step of 0.001 millisecond were around 5% over a one-second simulation period, as illustrated in Extended Data Figure 1. These errors tend to increase for longer simulation periods:., for data assimilation with fMRI data having a period of 0.8 seconds in the DB, the current level of precision is deemed acceptable. As shown in Supplementary Table 4, our GPU-based DB, by simulating spiking neuronal networks with 2, 3, 4, 5, and 10 million neurons on a single GPU, achieved real-time factors approximately 10 times smaller than those of the CPU-based tool NEST6 (v 3.7.0) on a single CPU under the same hardware and software environment (see Supplementary Section 5).

*Digital Brain in the resting state*

We first introduced the construction of the DB in the resting state using the neuronal network structure and data assimilation framework described above. We designed each voxel in the cortex as a six-layer micro-column structure[16,21,22], excluding Layer I (L1) and merging Layer II with Layer III into one layer (L2/3). During the assimilation process, sequential tracking of BOLD signals by tuning maximum synaptic conductance parameters was performed. The time series of estimation for hyperparameters for each voxel converged within 3-4 repetition times (TRs, meaning the period of the BOLD time series) of the observable resting-state fMRI signals, while the simulated BOLD signals presented a short delay compared to the experiment signals. To eliminate this delay, we set 2 TRs time-lag when computing the correlation between the empirical and simulated time series (see Supplementary Section 9) to maximize the average correlation in the resting-state experiment, which is fixed for the following task experiment. Then, the averaged correlation coefficients of the time-series among all voxels between the assimilated DB and its experimental counterpart reached 0.93 (Figure 2c). Importantly, the DB could reproduce patterns of spontaneous neuronal activity (Figure 2d). Certain subcortical regions (caudate and hippocampus) exhibited oscillatory states that were predominantly driven by excitatory activity within the beta-frequency bands (Figure 2e).

*Digital Brain in action*

Based on digital brain simulation and the HMDA method, we conducted a step towards establishing methodologies capable of assimilating the model with the experimental data in perceptual and cognitive tasks. Here, we outlined how visual evaluation was implemented in the DB, following the pipeline described in the previous work[23], there is currently no suitable encoding and decoding model to reproduce this intricate process in the field of biological modelling. To address this challenge, the DB decodes the input signal of the primary sensory region by assimilating the empirical BOLD signal. This decoded signal is then injected into the model to simulate the propagation of the signal within the best-fit resting-state configuration of the brain (Figure 2f). We could then obtain the simulated signals of the whole brain. A visual evaluation task[23] was performed by the subject, one of the authors in the present paper. For each trial, after a 3-second figure cue, the participant was asked to evaluate how pleasant they felt to the real-world stimulus on a Likert scale (ranging from 0 to 10) within 4 seconds (Figure 2g). In this task, total 30 trials were assessed.

For the specific task, we employed the HMDA algorithm on the perceptual region-of-interest (ROI) -referred to as the "input" brain region-at the voxel level to decode the sensory stimulus. The estimated hyperparameters of the voxels in the perceptive ROI served as injection currents (see Methods). The parameters of other voxels were taken from the DB of the resting state. The Calcarine (CAL) in the AAL template (See Supplementary Table 3) was used as the input region for decoding the perceptual inputs in the context of data assimilation. meaning, the hyperparameters of the injection currents of all voxels in CAL were estimated by the BOLD signals. The Pearson correlation coefficients (PCC) between assimilated and the real BOLD signals were computed to measure the similarity between the DB and the biological brain. The PCCs reached over0.98 with a time lag of 2 second in the perceptual input region. The average PCC over the braon was 0.53 (Figure 2h) but only 0.37 and 0.35 within the subcortex and brainstem, respectively, while can reach 0.79 in the cerebellum (Extended Data Figure 2 left). These region-level PCCs were highly associated with the number of connections between these regions and the perceptual input region ($r = 0.678$, $p = 3.8E-11$, Extended Data Figure 2 right), suggested that the propagation of the signals in the DB is possibly driven by the brain architecture.

We then used the DB to predict the scores in the visual task. To achieve this, brain activation of the stimulus cue for each trial was first assessed, using the BOLD signals from both the real and digital task brains, by a general linear model. In this model, the BOLD signals were deemed as the predictors, and regressors for modelling each trail were established by convolving the onsets of the corresponding experimental condition (the binary task design matrix) with SPM's canonical hemodynamic response function (HRF), along with six head motion parameters set as the additional covariate regressors. Thus, the patterns of brain activation were comparable between the DB and the biological brain (Figure 2i left). Second, we trained a linear model with the biological brain activations during simulations, as the response variables and the real scores of the emotional pictures as the predictors. Finally, we predicted how the subject rated real-world pictures with the brain activations from the DB and the sparse coefficient vector obtained via LASSO (Figure 2i middle). Notably, the predicted scores from the DB were consistent with the actual scores ($r = 0.575$, $p < 0.001$, Figure 2i right).

## Discussion

It is essential to establish the DB as a network model at neuron and synaptic levels, because the neuron is the basic unit for brain function and the interaction between neurons is at the fundamental level for understanding brain function[24]. In addition, the scale of the neuronal networks, with brain architecture, affects its performance in both resting-state and tasks[25].

The DB opens up the possibility of testing various experiments in simulating human brains and will be open for scientists all over the world to perform 'dry' experiments and test hypotheses towards answering important questions in neuroscience, brain medicine, and artificial general intelligence. While other neuromorphic systems such as SpiNNaker, SpiNNaker 2, IBM TrueNorth and Intel Loihi may have the potential to support large-scale simulations and assimilations, there are challenges for large-scale human brain simulations and assimilations. They arises not only from insufficient hardware resources but also from the heterogenous long-range connections between neurons and synapses.

We would also like to note some limitations. First, we only have limited biological data to explore the network architecture. For example, DWI data is directionless but the brain network topologies are directed[26]. Second, the bandwidth limitations of both communication and memory access prevent achieving the biological scale and precisions of brain simulation. For example, the average synaptic degree of each neuron is set to up to 1000, though this is widely known to be even as large as 10,000 in the human brain[27], which has been demonstrated to possess significant influence on network dynamics[28]. Third, the setup of the neuronal network model has much space to be improved, because many details of synaptic densities, microarchitectures, neurobiological details of the synaptic environment noise etc have been simplified. For instance, the current version did not consider the following factors: the synaptic delays[29], diverse types of temporal kernel of the synapse model [30], a dozen biologically important synapse receptors d and ion-based channels, adaptive threshold and, more importantly, the self-organized dynamic weights related to learning, which could essentially affect the dynamics of neuronal networks[31]. These omissions may lead to a disagreement of neural dynamics between the DB model and neurophysiological/clinical data. Finally, the performance of fitting the BOLD signals at the voxel resolution is limited to validate the neuroscientific value of the DB, which needs further investigations.

To warrant the applicability of the DB, we need to work on a range of more fundamental infrastructure aspects. For example, one can replace the hippocampus region with the recently developed more realistic, biophysical-based hippocampus model[32] as a plug-in subnetwork in the DB model. In additionthe interaction between the digital model and its biological counterpart is an important issue in establishing a digital twin brain with scientific implications for studies in neuroscience, brain medicine, and brain-inspired intelligence. In such a way, the integrated, gradually improved DB should be a better digital tool to deepen the understanding of our brain. Finally, the current model needs a considerable amount of computational power to run, and a hardware implementation should be one of the future aims.

## Methods

*Constructing the Digital Brain model*

The computational basis of the DB is composed of two components: the basic computing units and the network architecture. The basic computing units of the DB are neurons and synapses, and action potentials (that is, spikes) transmitted between neurons. The network architecture gives the synaptic interactions between neurons in the form of a directed multiplex graph (see Supplementary Section 2 and Extended Data Tables 2,3).

*Simulating Digital Brain on GPUs*

We simulated a neuronal network with a computational neuron model and synaptic network model, by the high-performance-computing (HPC) system of GPUs, due to their power of parallel computation. The simulation of the neuronal network was composed of two components: the spike integrals of the synaptic input spike trains over membrane potentials and spike communications that transferred spike trains between neurons.

The computational neuron model of the DB could generally be a nonlinear operator from a set of input synaptic spike trains to an output axon spike train. In this work, the neurons were based on the leaky integrate-and-fire (LIF) model[33]: A capacitance-voltage (CV) equation described the membrane potential of neuron *i* when it is less than a given voltage threshold $V_{th,i}$:

$$C_i \frac{dV_i}{dt} = -g_{L,i}(V_i - V_L) + \sum_u I_{syn,i} + I_{bg,i} + I_{ext,i},\ V_i < V_{th,i}$$

where $C_i$ is the capacitance of the neuron membrane, $g_{L,i}$ is the leakage conductance, $V_L$ is the leakage voltage, $I_{bg,i}$ is the background current from the environment, serving as noises, and $I_{ext,i}$ is the external stimulus, serving as the external injection input for the task. It should be noted that the $I_{ext,i}$ are the parameters, which are independently sampled following a Gamma distribution with certain hyperparameters estimated. Four synapse types (i.e., AMPA, NMDA $GABA_A$, and $GABA_B$) are considered in this model. The background current is given by independent Ornstein Uhlenbeck (OU) processes and described as follows:

$$\tau_{bg} dI_{bg,i} = (\mu_{bg} - I_{bg,i})dt + \sqrt{2\tau_{bg}}\sigma_{bg} dW_{i,t} \quad (1)$$

where $\tau_{\mathrm{bg}}$, $\mu_{\mathrm{bg}}$, and $\sigma_{\mathrm{bg}}$ represent the s time-scale constants, the stationary mean, and the standard variance of the process, respectively, which areuniform among neurons, and $W_{i,t}$ is a Brownian motion. When $V_i = V_{th,i}$ at $t = t_n^i$, the neurons registere a spike at time point $n$ and the membrane potential is reset at $V_{rest}$ during a refractory period. After that, $V_i$ is governed by the CV equation again. In the case that neuron $j$ was connected to neuron $i$, an exponential temporal convolution was established:

$$\begin{aligned} I_{u,i} &= g_{u,i}(V_u - V_i)J_{u,i} \\ \frac{dJ_{u,i}}{dt} &= -\frac{J_{u,i}}{\tau_i^u} + \sum_{k,j} w_{i,j}^u \delta(t - t_k^j) \end{aligned} \quad (2)$$

where $g_{u,i}$ is the maximal conductance of synapse type $u$ of neuron $i$, $V_u$ is the voltage of synapse type $u$, $\tau_i^u$ is the time-scale constants of synapse type $u$ of neuron $i$, $w_{i,j}^n$ is the connection weights from neuron $j$ to $i$ of synapse type $u$, $\delta(\cdot)$ is the Dirac-delta function and $t_k^j$ is the time point of the $k$-th spike of neuron $j$.

All numerical calculations about neurons use a first-order Euler-Maruyama method with an iterative time-step of 1 msec or 0.1 msec according to application. The time step of spike communication is fixed as 1 msec, as achieved by the Message-Passing Interface (MPI) [34] with communication cables which can be regarded as the axons connecting the neurons.

There are two major approaches to accelerating the simulation performance. One is the parallelization of the computation and communication threads. The entire computational cycle begins with updating the membrane potential equation by the computational thread, and ends with completion of a maximum synaptic conductance calculation, upon receiving spikes, thus completing one full iteration. All kernel operations synchronize once the computational thread finishes, ensuring that all nodes have completed one full iteration before initiating the next iteration's computational kernel. This synchronization operation guarantees that messages from the previous iteration's communication have been received and processed, avoiding information misalignment. This significantly decreases the total simulation duration (see Supplementary Section 4). However, it also means that the overall computational time depends on the slowest node's computation, thus the load balancing of the network determines the overall efficiency of the model simulation.

The other approach is the optimized partition algorithm to assign voxels of neurons on the GPUs and the two-level routing method to balance the computation and communication. The main time cost of the DB simulation depended on the communication of synaptic spikes between neurons and memory access. The neurons of the DB model were allocated into the GPUs with a partitioning algorithm and a two-level routing method. Specifically, instead of allocating neurons to multiple GPUs according to their biological layout, we applied a partitioning algorithm to assign a set of neurons with strong communication between each other to the same GPU to take advantage of the faster communication speed within a GPU. According to the data flow generated by neurons and synapses and the performance of GPU, the partitioning algorithm can balance the traffic load across all processes, thereby effectively supporting such large-scale human brain simulations and assimilation. We further split multiple GPUs into different groups to minimize the amount of traffic between GPUs in different groups. An iterative greedy algorithm was employed to assign GPUs to different groups to ensure optimal traffic flow within a group and minimize coupling between groups. Within the same group, GPUs are interconnected directly for communication. If a GPU sends data to a GPU in another group, it initially passes the data to a bridge GPU within its group, which relays the data to the target GPU in the other group. Compared to the original one-level routing communication, we succeeded in reducing the total number of source and/or destination message counts to a magnitude order of the square root of the original count [35].

Due to the necessity of initiating a separate thread for each message count, it was possible to mitigate the time expenditure associated with thread initiation throughout the entire system by reducing the number of counts. It is noteworthy that the simulation and assimilation of higher synaptic densities depend on the data flow generated by neurons and synapses and the performance of the hardware. Based on extensive experimentation, we found that under the same hardware conditions, if synaptic densities increase by 10 times, the hardware requirements increase by approximately 9 times. Even with increased synaptic density, our partitioning strategy can still effectively balance the load and data flow to keep the simulation time from growing proportionally[13,14,35]. We completed simulation experiments of a human brain network of 86 billion neurons and 47.8 trillion synapses on a cluster of heterogeneous computing nodes. The main steps included: (1) generating neuron attribute tables for the excitatory/inhibitory neurons as well as their parameters, and connection tables describing the labels of identities (IDs) of neurons that are connected to this neuron, (2) running the pre-simulation process to load data, mentioned in the first step, deploy neurons by their IDs to each GPU, and assign the simulation data recording setup, (3) initializing the simulator, (4) running the simulator to complete simulation accompanied with simulating data recording.

*Statistical inference of the Digital Brain by hierarchical mesoscopic data assimilation (HMDA)*

The canonical Bayesian inference framework implied that the number of parameters involved in the neuronal network model was possibly much larger than the number of data points, which essentially leads to overdetermination if conducting this sort of Bayesian inference. Alternatively, we considered the hierarchical Bayesian inference by introducing hyperparameter $\vartheta$ to describe the distribution of parameters (Extended Data Figure 3), which contained the important neurophysiological information of these parameters:

$$P(\vartheta \mid y_t) \propto \int P(y_t \mid x_t, \theta) P(x_t \mid \theta) P(\theta \mid \vartheta) P(\vartheta) d[x_t] d\theta .$$

Here, $y_t$ stands for the observation, e.g., the BOLD signals in this study, $x_t$ for the internal states, e.g., the membrane potentials, synaptic currents, and the states of Balloon-Windkessel model, $\theta$ for the parameters, e.g. the maximum conductance and injection currents, and $\vartheta$ for their hyperparameters(see Supplementary Section 6). Hyperparameters could greatly reduce the number of variables to be inferred, thus solving the over-fittingness of the statistical inference. However, it was essential to identify the hyperparameters as well as the parametric distribution of the parameters, towards balancing between overfitting and preciseness of models.

We considered each voxel or layer of micro-column as a sub-unit and assumed that the maximum synaptic conductance parameters of the same type of neurons within the same sub-unit shared the same hyperparameters. The time series of BOLD signals from each voxel were taken as the observations of our neuronal network model, which were generated by the Balloon-Windkessel model based on the neural activity quantified with the spike rate of a pool of neurons[19] (Extended Data Table 4).

The computation complexity of each iteration depends on the dimension of the observed data. The distributed Kalman filter is an efficient way to handle high-dimensional observations 错误!未找到引用源。. In our case, both experimental observations with high resolution and a limited number of time points would cause both computation complexity and ill-posedness. Herein, we utilized this idea in the diffusion ensemble Kalman filter (EnKF) by taking the BOLD signal of each voxel as an observer and establishing an independent EnKF (Supplementary Algorithm 3). Correction of the states, parameters, and hyperparameters of each voxel was weighted average between EnKF of itself and EnKF from other voxels 错误!未找到引用源。. A fusion coefficient $\gamma$ was introduced to balance the fusion of corrections from itself and all others (see Supplementary Section 6).

*Statistical inferring Digital Brain in resting state*

Although HMDA could obtain all hidden states and parameters contained in the model, we assumed that the neuronal network was driven solely by the maximum synaptic conductance. Therefore, after estimating the hyperparameter of each sub-unit with an observation signal, we re-simulated this model by assigning the maximum synaptic conductance to each neuron according to the hyperparameter series ($g_u$). It should be noted that the assimilated hyperparameters have been employed in the DB model with the same MRI-based structure but at different scales, including different numbers of neurons and different average synaptic connection degrees Simulating the DB is carried out by sampling the parameters by the corresponding estimated hyperparameters. For instance, the maximum synaptic conductance parameters of the network with different average degrees ($d$) are simply read as

$$g_{u,i,D} = g_{u,i,d} \frac{D}{d}. \qquad (3)$$

In the initial assimilation of the DB, we used a neuronal network of 200 million neurons with an average of 100 synaptic connections per neuron and then assimilated the rest-state BOLD signals with this model.

*Statistical inferring Digital Brain in task*

In the task work, the hyperparameters of the maximum synaptic conductance $g_{u,i}$ were used to statistically infer the model from the resting-state fMRI data and those of the external current $I_{ext,i}$ were used to statistically infer the model with task fMRI data. According to the specific task, we estimated the hyperparameters of the input currents $I_{ext,i}$ to the neurons in the voxels that are located in the perceptive ROI defined by the task fMRI data in the whole brain neuronal network. We employed the HMDA on the perceptive ROI (refer to "input" brain region) at the voxel level to decode the sensory stimulus and the assimilated hyperparameters of the voxels in the perceptive ROI will serve as the injection currents. For instance, in the visual evaluation task, the calcarine (CAL) has been included as the perceptive input, which is responsible for the primary visual information processing.

**Data Availability**

The datasets that were used to establish and validate to the Digital Brain in this Brief Communication are publicly available at https://github.com/DTB-consortium/Digital_twin_brain-open. Source data for all figures are available with this Brief Communication. All requests for further information of the datasets should be addressed to and fulfilled by our group: the DTB Consortium, Institute of Science and Technology for Brain-Inspired Intelligence, Fudan University, dtb.fudan@gmail.com.

**Code Availability**

Custom codes in Pytorch for PC and C++ for the HPC systems of our DB platform can be accessed on our GitHub profile at https://github.com/DTB-consortium/Digital_twin_brain-open. We must note that our code strongly depends on the hardware and software conditions of our HPC system and may not run on other HPC systems. All requests for further information of code should be addressed to our group: the DTB Consortium, Institute of Science and Technology for Brain-Inspired Intelligence, Fudan University, dtb.fudan@gmail.com.

**Acknowledgments**

We thank Professor Trevor W. Robbins, Professor Edmund Rolls, Professor Karl Friston, and Professor David Waxman for their helpful comments on the manuscript. The simulation was supported by Advanced Computing East China Sub-Centre. This work received support from the following sources: STI2030-Major Projects 2021ZD0200204, Shanghai Municipal Science and Technology Major Project (No. 2018SHZDZX01), ZJ Lab, Shanghai Centre for Brain Science and Brain-Inspired Technology, the 111 Project (No. B18015) and the National Natural Science Foundation of China (No. 62072111).

**Author Contribution Statement**

W.L., Q.Z. and J.F. conceptualized the study; W.L., Q.Z. and J.F. designed the analytic approach; W.L., S.X, L.Z, J.W, X.D. and DTB Consortium finished investigation; S.X., L.Z. and DTB Consortium helped visualization; N.X. and J.F. acquired funding supports; Q.Z. and J.F. helped project administration; W.L., S.X., L.Z., J.W. and DTB Consortium wrote the manuscript; W.L, S. X., L. Z., J.W., X.D. and J.F. revised the first draft; All authors critically revised the manuscript.

**Competing Interests Statement**

The authors declare no competing interests.

Figures

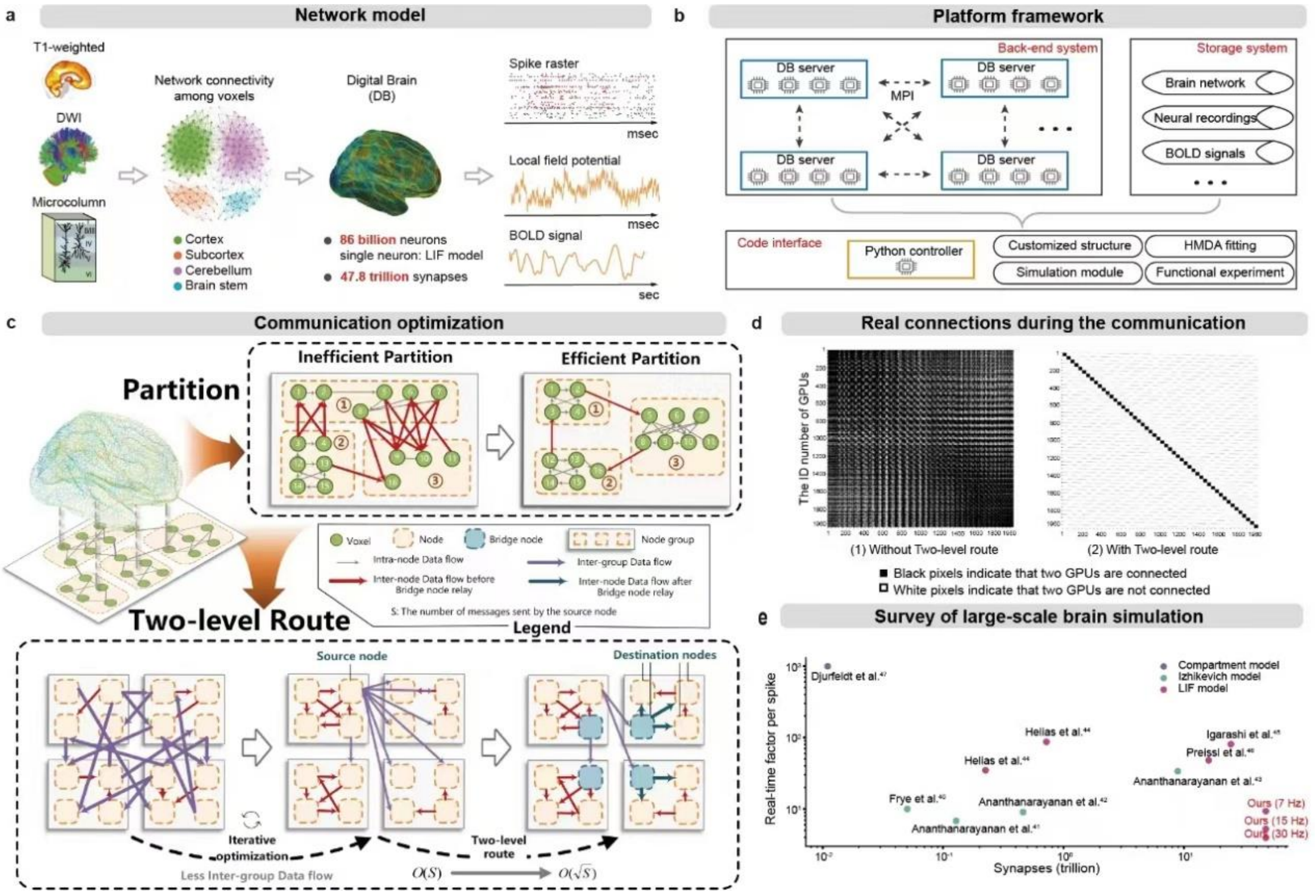

**Figure 1. Schematic representation of the Digital Brain (DB) workflow and simulation performance.** (a) The DB is constructed with different numbers of spiking neurons to simulate and imitate the brain activity. Multi-modality MRI data including DWI and T1-weighted MRI, and a microcolumn connection map were used to construct a probabilistic connection network (left column). The whole brain was divided into four different parts (i.e., cortex, subcortex, brainstem, and cerebellum). Within each voxel in the cortex, the spiking neurons were connected following the microcolumn structure. The neurons within each voxel in the other three parts were connected following certain without considering the microcolumn structure (left middle column). The DB can generate diverse modal function data, for example, spike raster, local field potentials, and BOLD signal (right column). (b) The Client-Server framework of the DB software. The DB configured for Client-Server deployment consists of the back-end system, storage system, and interface, with brain model instances hosted on a server accessible to client units. (c) Illustration of the communication optimization process. The optimized partition algorithm assigns voxels of neurons on the GPUs that optimize data traffic by minimizing graphics processing unit (GPU) connections, and the two-level routing method, further balances the communication traffic and parallelizes the communication processing by introducing the bridge node to transfer the communication between two GPUs with a high load of traffic (see Methods and Supplementary Section 4 for details). Communication optimization can reduce the number of neuronal connections from each source (S) to each destination to the square root of their original counts[13]. (d) Real connections during the communication of an HPC with 2000 GPUs. The horizontal and vertical axes represent the ID number of GPUs. Black pixels indicate that two GPUs are connected, while white pixels indicate that two GPUs are not connected. Left panel: The connections between GPUs without the two-level routing method; Right panel: The connections between GPUs without the two-level routing method. (e) Performance comparison of recent large-scale neuronal network simulation according to the number of synapses and the real-time factor per spike.

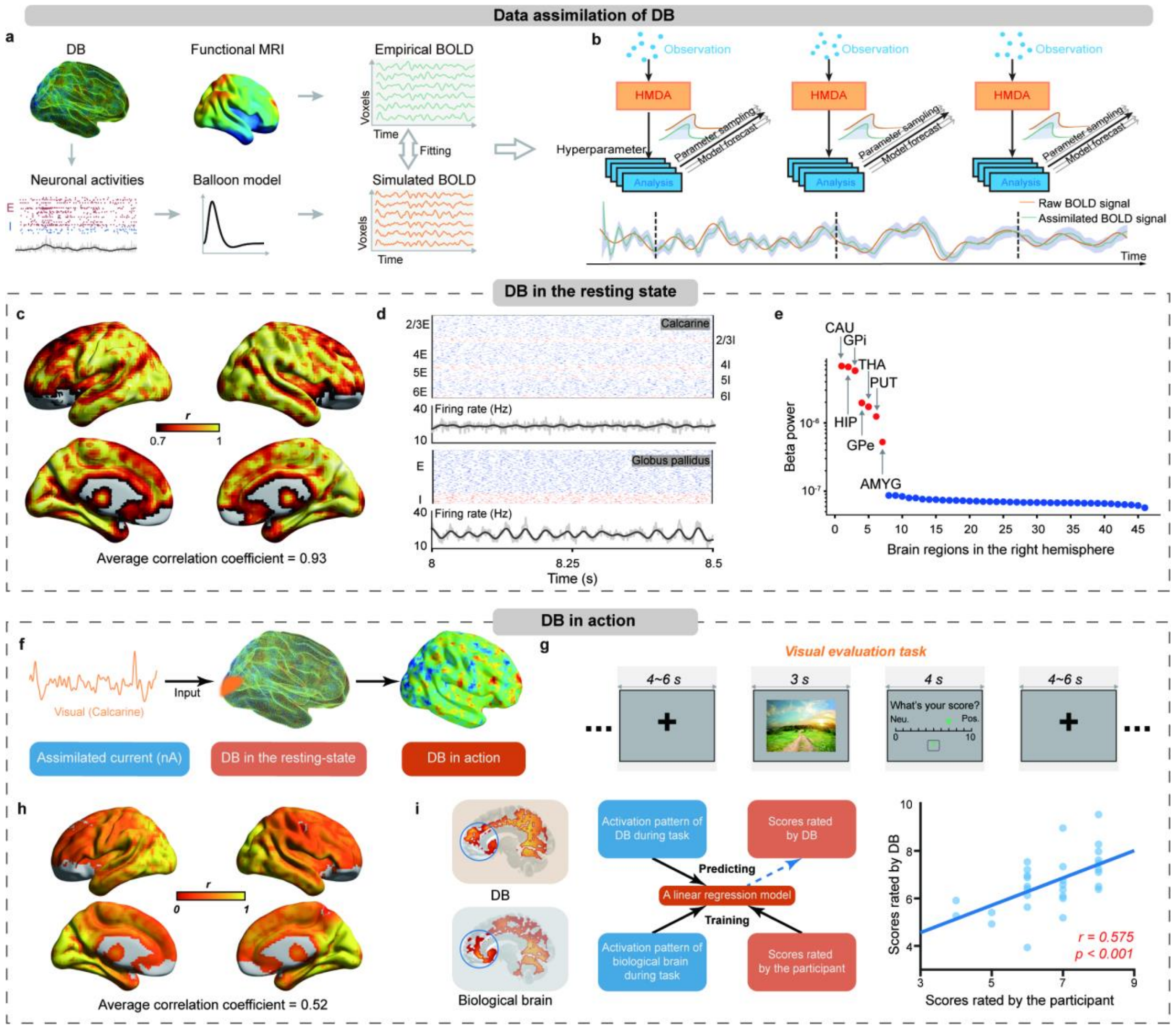

**Figure 2. Digital Brain in the resting state and in action.** (a) The Digital Brain (DB) uses a LIF neuron model with four synapses (AMPA, NMDA, $GABA_A$, and $GABA_B$) to simulate spike activity. Firing rates, derived from spike counts over a sliding window, are input into the Balloon-Windkessel model to generate simulated BOLD time series. Maximum AMPA conductance (resting state) and postsynaptic currents (action) were estimated to match empirical BOLD signals (see (b)). (b) The HMDA method combines diffusion EnKF and hierarchical Bayesian inference to estimate DB parameters by comparing raw and assimilated BOLD signals at marked time points (dashed lines). (c) Pearson correlations between empirical and assimilated BOLD signals at voxel resolution, averaging 0.93 across the brain. (d) Spike raster and firing rate plots for specific cortical (Calcarine) and subcortical (Globus Pallidus) voxels, with excitatory neurons in blue and inhibitory neurons in red. (e) Power spectrogram of beta rhythms in right hemisphere subcortical regions, highlighting areas with elevated beta power, including the Caudate Nucleus (CAU), Globus Pallidus Internus (GPi), Thalamus (THA), and others. (f) Workflow for the DB in action, beginning with the assimilation of BOLD signals from the primary visual cortex (orange) to estimate input postsynaptic

currents, which are injected into the DB for task-related fMRI data. (g) Experimental design for the visual evaluation task, where participants assess real-world stimuli on a Likert scale (0-10) after a 3-second cue, with inter-trial intervals randomly set between 4 and 6 seconds. (h) Voxel-wise Pearson correlations between empirical and assimilated BOLD signals, averaging 0.53 overall and approximately 0.98 in the primary visual region. (i) Prediction of evaluation scores in the visual task using simulated voxel-wise BOLD signals from the DB. The left panel shows activation pattern similarities between the DB and biological brain, marked by a blue circle. A linear regression model trained on biological activations and actual scores (middle panel) indicated a significant correlation (Pearson $r=0.575$, $p< 0.001$, $df = 29$). Tests were two-sided, with no multiple comparison adjustments, yielding a p-value of 3.6e-4, an effect size of 0.575, and a 95% CI of [0.435, 0.688] (right panel).

**The DTB Consortium**

Yubin Bao[1], Boyu Chen[2], Siming Chen[1], Zhongyu Chen[3], Fei Dai[3], Weiyang Ding[3], Xin Du[1,3], Jianfeng Feng[1,,3,4,5,6,7,8], Yubo Hou[2], Mingda Ji[3], Peng Ji[3], Chong Li[3], Chunhe Li[5], Xiaoyi Li[3], Yuhao Liu[3], Wenlian Lu[2,3,4,5], Zhihui Lv[1], Hengyuan Ma[8], Yang Qi[3], Edmund Rolls[3], He Wang[3], Huarui Wang[3], Jiexiang Wang[3,4], Shouyan Wang[3], Ziyi Wang[3], Yunman Xia[3], Shitong Xiang[3,4], Chao Xie[3], Xiangyang Xue[3], Leijun Ye[3,4], Longbin Zeng[3,4], Tianping Zeng[3], Chenfei Zhang[3], Jie Zhang[3,4], Nan Zhang[3], Wenyong Zhang[2], Yicong, Zhao[3], Qibao Zheng[3,4].

Affiliations:

[1] School of Computer Science, Fudan University, Shanghai, China

[2] School of Mathematical Sciences, Fudan University, Shanghai, China

[3] Institute of Science and Technology for Brain-Inspired Intelligence, Fudan University, Shanghai, China

[4] Key Laboratory of Computational Neuroscience and Brain-Inspired Intelligence (Fudan University), Ministry of Education, China

[5] Shanghai Centre for Mathematical Sciences, Fudan University, Shanghai, China

[6] Department of Computer Science, University of Warwick, Coventry, United Kingdom

[7] Zhangjiang Fudan International Innovation Centre, Shanghai, China[8] School of Data Science, Fudan University, Shanghai, China

# Extended Data

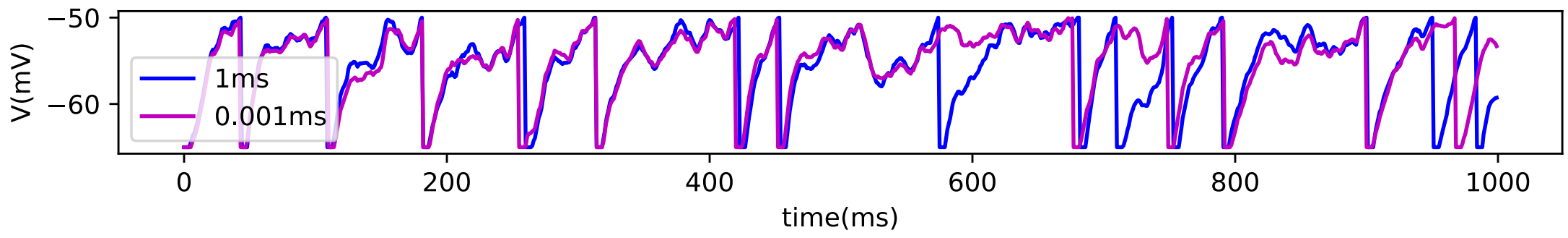

**Extended Figure 1. Preciseness of Simulation of the Euler-Maruyama method.**
The comparison of the membrane potentials in a LIF neuronal network of 1000 neurons by the Euler-Maruyama method with time step 1 msec (blue plot) and 0.001 msec (purple plot).

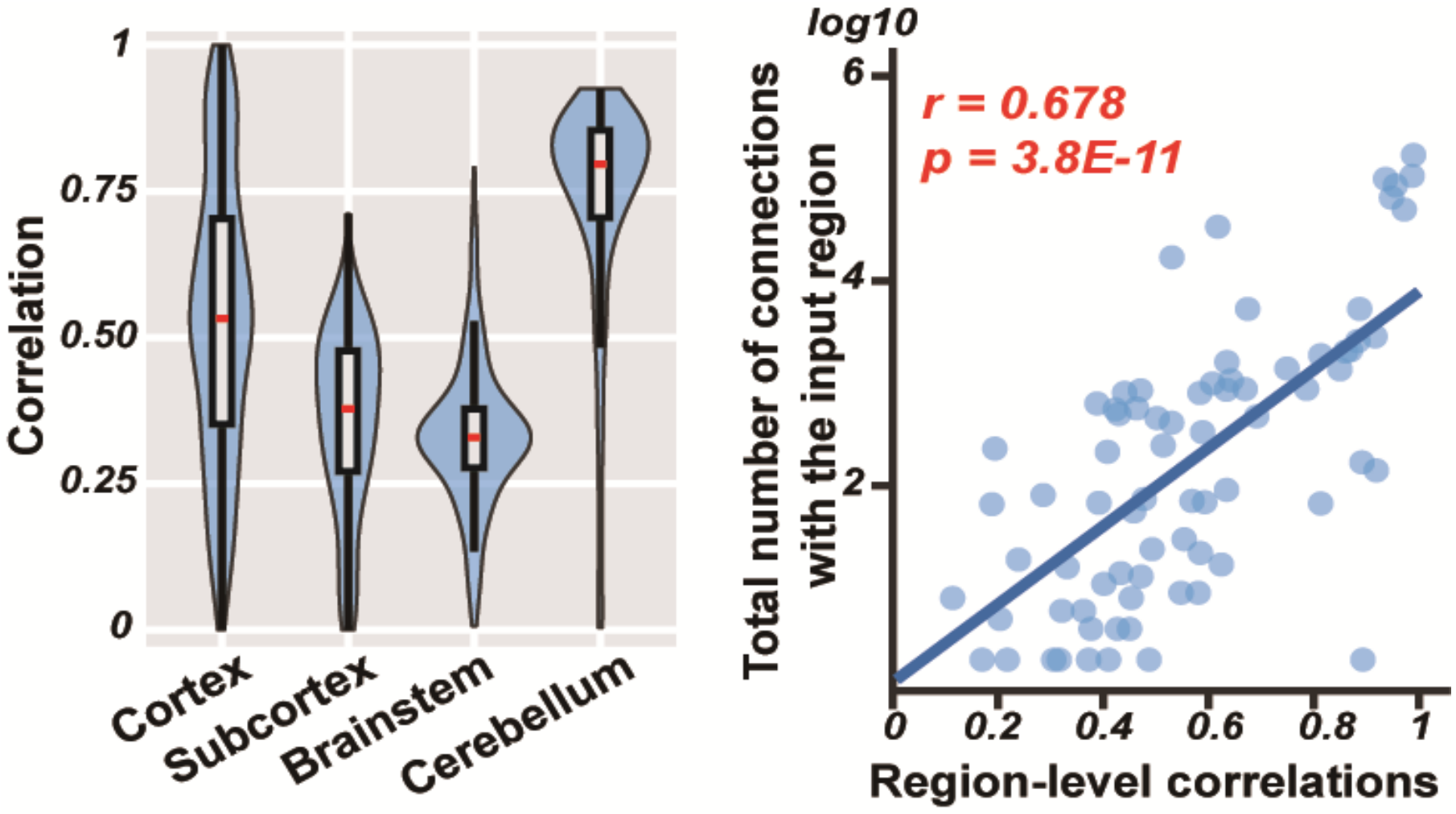

**Extended Data Figure 2. Digital brain (DB) in action (visual evaluation task).** The distributions of Pearson correlations between the empirical and assimilated BOLD signals at voxel-level in different brain parts were evaluated in the left panel. The box represents the interquartile range (IQR) from the 25th to the 75th percentile, with the median indicated by a line within the box. Whiskers extend to the minimum and maximum values, excluding outliers, providing a comprehensive overview of the data distribution. In the right panel, region-level correlations showed a positive association with the strength of structural connections (obtained from DWI) between the corresponding brain region to the input region (Pearson r=0.678, p=3.8e-11, df=120). Tests were two-sided, with no multiple comparison adjustments, yielding an effect size of 0.678, and a 95% CI of [0.625, 0.724] (right panel).

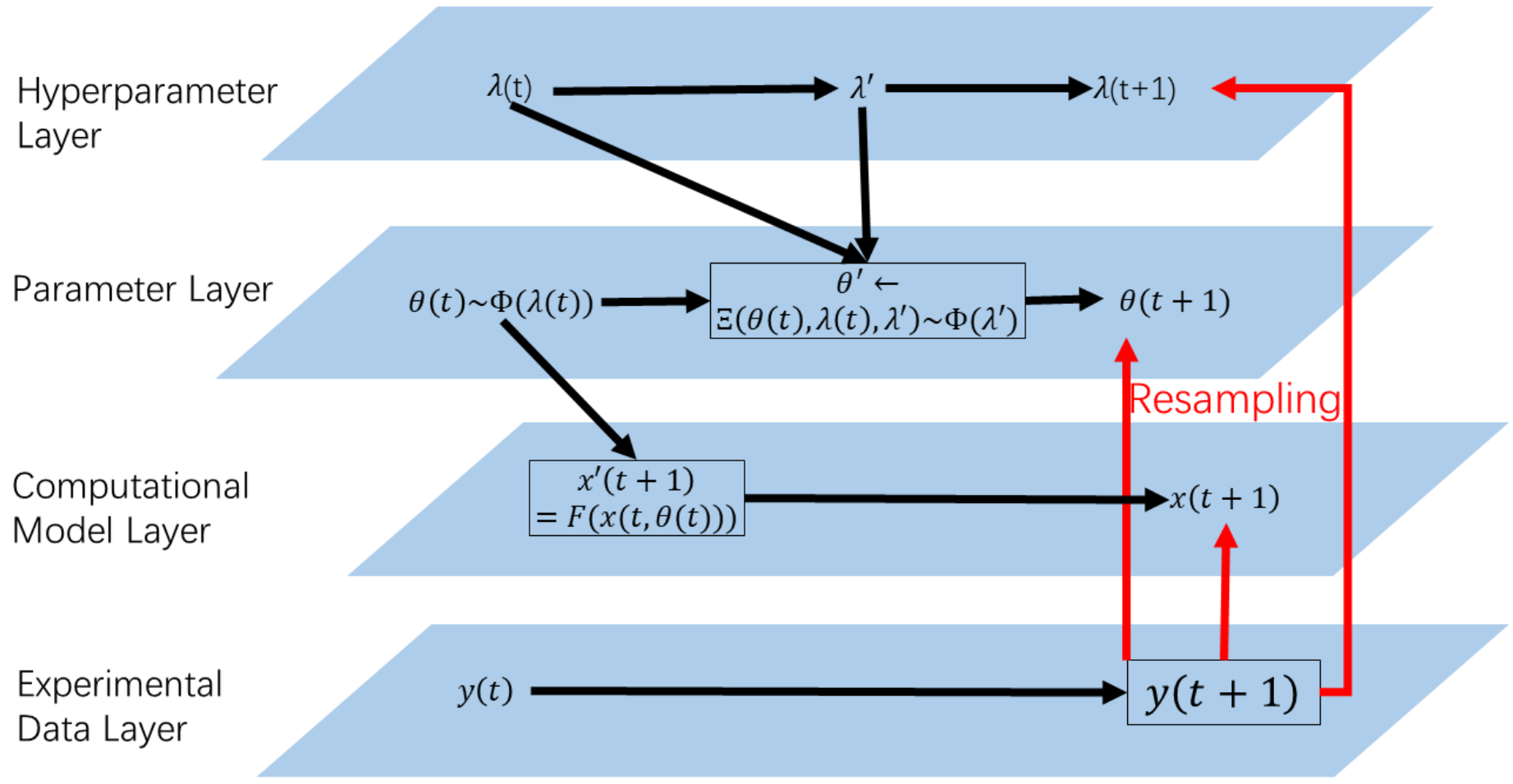

**Extended Data Figure. 3. Framework of Hierarchical Bayesian Inference.**

| Extended Data Table 1: Statistical Information of the Brain Network | | | | |
|---|---|---|---|---|
| Regions | Cortex | Sub-cortex | Brainstem | Cerebellum |
| Neuron Nos per voxel (mean ± std) | 834369 ± 127903 | 290432 ± 97269 | 110796 ± 46885 | 42528898 ± 8384971 |
| Input connection Nos per neuron | 1000 | 1000 | 100 | 100 |
| Output connection Nos per neuron (mean ± std) | 1004 ± 351 | 952 ± 260 | 202 ± 185 | 103 ± 34 |
| Maximum output connection Nos per neuron | 5017 | 3612 | 1542 | 833 |
| Minimum output connection Nos per neuron | 714 | 760 | 44 | 87 |
| Voxel-voxel connection Nos (mean ± std) | 170 ± 213 | 250 ± 283 | 147 ± 109 | 108 ± 105 |
| Total Number of Voxels | 19570 | 2191 | 480 | 1622 |
| Structure of Voxels | Microcolumns | Voxels | Voxels | Voxels |
| Total Number of Neurons | 16.33 billion | 0.64 billion | 0.05 billion | 68.98 billion |

| Extended Data Table 2: The summary of the Digital Brain model | |
|---|---|
| Brain part | 4 brain parts consist of the cortex, sub-cortex, brainstem, and cerebellum. |
| Topology | Voxel-wise DTI is normalized (row normalization) as the connection probability. |
| Populations | The cortical voxel is described as microcolumn (8 populations for each) and the others are modelled as a unified voxel structure (Exc. and Inh. 2populations). |
| Connectivity | Population-specific but otherwise random. |
| Neuron model | Leaky integrate-and-fire (LIF). |
| Synapse model | Conductance chemical synapse (4 types), exponential postsynaptic currents. AMPA and NMDA are modelled in excitatory projection; GABAa and GABAb are modelled in inhibitory projection. |
| Background input | independent homogeneous Poisson spike trains or OU-type external current. |
| Neuronal size | The proportion of neurons distributed in 4 brain parts is as $N_{cortex}: N_{cerebellum}: (N_{brainstem} + N_{subcortex}) = 19\%: 80\%: 1\%$. We then assign number of neurons in each voxel is proportional to their corresponding gray volume and the ratio of the number of excitation neurons over that of inhibitory neurons of each layer/voxel equals 4： 1. |
| Input | |
| Poisson spike | Each synapse is accompanied by 10hz independent Poisson spikes. |
| OU current | The background current is given by independent Ornstein Uhlenbeck (OU) processes and described as: $\tau_{ext} dI_{ext,i} = (\mu_{ext} - I_{ext,i})\mathrm{dt} + \sqrt{2\tau_{\mathrm{ext}}}\sigma_{ext} dW_t$. The parameters are the same among neurons ($\mu_{ext} = 0.66, \sigma_{ext} = 0.12, \tau_{ext} = 10$). |
| Connectivity | |
| Type | Source and target neurons are drawn randomly without replacement (not allowing autapses and multapses) with population-specific connection probabilities. |
| Population-specific connection probability. | The count of inner and outer synaptic edges of each population is decided by the DTI and micro-circuit (micro-column or voxel structure). The detailed micro-circuits inside the voxel of the cortex, subcortex, brainstem, and cerebellum are listed in Table 2. |
| In-degree | The average in-degree of neurons in a voxel equals a constant. We set 1000 in cortex and subcortex and 100 for others. |
| Weight | fixed, drawn from uniform distribution $U[0, 1]$. |

**Extended Data Table 3: Average number of synapses received by individual neurons in each cortical layer**

| postsynaptic \ presynaptic | i1 | e2/3 | i2/3 | e4 | i4 | e5 | i5 | e6 | i6 | CC |
|---|---|---|---|---|---|---|---|---|---|---|
| e1 | 1323 | 823 | 200 | 15 | 1 | 9 | | | | 10133 |
| i1 | 901 | 560 | 149 | 10 | 1 | 6 | | | | 6899 |
| e2/3 | 133 | 3554 | 804 | 881 | 45 | 431 | | 136 | | 1020 |
| i2/3 | 52 | 1778 | 532 | 456 | 29 | 217 | | 69 | | 396 |
| e4 | 27 | 417 | 84 | 1070 | 782 | 79 | 8 | 1686 | | 1489 |
| i4 | | 168 | 41 | 628 | 538 | 36 | | 1028 | | 790 |
| e5 | 147 | 2550 | 176 | 765 | 99 | 621 | 596 | 363 | | 1591 |
| i5 | | 1357 | 76 | 380 | 32 | 375 | 403 | 129 | | 214 |
| e6 | 2 | 643 | 46 | 549 | 196 | 327 | 126 | 925 | 597 | 2609 |
| i6 | | 80 | 8 | 92 | 3 | 159 | 11 | 76 | 499 | 1794 |

e1,2/3,4,5,6 and i1,2/3,4,5,6 represented the excitation and inhibition neuron population of the layer 1,2/3,4,5,6 respectively.

**Extended Data Table 4: Details of the Balloon-Windkessel model**

| Equations | Notations |
|---|---|
| $\dot{s}_i = z_i - \kappa_i s_i - \gamma_i (f_i - 1)$ | $z_i$ the neural activity; $s_i$ vasodilatory signal; *i* the label of voxel |
| $\dot{f}_i = s_i$ | Inflow |
| $\tau_i \dot{v}_i = f_i - v_i^{\frac{1}{\alpha}}$ | Blood volume; $f_{out,i} = v_i^{\frac{1}{\alpha}}$ |
| $\tau_i \dot{q}_i = \frac{f_i E(f_i, \rho_i)}{\rho_i} - v_i^{\frac{1}{\alpha}} q_i / v_i$ | Deoxyhemoglobin content, $E(f, \rho) = 1 - (1 - \rho)^{\frac{1}{f}}$ |
| $y_i = V_0 [k_1 (1 - q_i) + k_2 \left(1 - \frac{q_i}{v_i}\right) + k_3 (1 - v_i)]$ | Bold signal |

# Supplementary Information

## Section 1: Biological data acquisition and pre-processing for Digital Brain

In this work, we scanned multimodal MRI from the corresponding author of this paper, Jianfeng Feng. All neuroimaging was performed on a 3 Tesla MR scanner (Siemens Magnetom Prisma, Erlangen, Germany) at the Zhangjiang International Brain Imaging Centre in Shanghai, using a 64-channel head array coil. High-resolution T1-weighted (T1w) images were acquired using a 3D Magnetization-Prepared RApid Gradient Echo (3D-MPRAGE) sequence (TR/TE = 3000/2.5 ms, TI = 1100 ms, flip angle = 7°, FOV: 320*320*240, Voxel size: 0.8*0.8*0.8 $mm^3$). Multi-shelled diffusion-weighted images (DWI) were acquired using a single-shot spin-echo planar imaging (EPI) sequence (TR = 3200 ms, TE = 82 ms, FOV: 140*140*92, Voxel size: 1.5*1.5*1.5 $mm^3$, multiband factor = 4, phase encoding: anterior to posterior) with two b-values of 1500 s/$mm^2$ (30 diffusion directions) and 3000 s/$mm^2$ (60 diffusion directions), in which B0 images were interleaved in every 6 volumes. B0 images with the same DWI protocol using an opposite polarity (i.e., phase encoding from posterior to anterior) were also acquired. fMRI data were acquired using a gradient echo-planar imaging (EPI) sequence (TR = 800 ms, TE = 37 ms, flip angle = 52°, FOV: 104*104*72, Voxel size: 2*2*2 $mm^3$, multiband factor = 8, phase encoding: anterior to posterior). The resting-state fMRI scan consisted of 400 contiguous EPI volumes and task-based fMRI scans consisted of 350 and 570 EPI volumes in visual and auditory evaluation tasks, respectively.

The voxel-based morphometry (VBM) of T1w images was pre-processed by the VBM8 toolbox in the Statistical Parametric Mapping package (SPM, http://www.fil.ion.ucl.ac.uk/spm). Briefly, the gray matter image was segmented and normalized to Montreal Neurological Institute (MNI) space by a nonlinear registration. Finally, the normalized image was smoothed with a full-width at half-maximum (FWHM) 8-mm Gaussian kernel and resampled at a resolution of 3*3*3 $mm^3$. The DWI was denoised, corrected for Gibbs ringing artifacts, head motion, eddy current, and tissue susceptibility-induced off-resonance geometric distortions with the reversed phase encoding B0 images and adjusted biased field by FSL software (version 6.0.4, http://www.fmrib.ox.uk/fsl) and MRtrix3 (http://www.mrtrix.org) [13]. For the fibre tracking, we generated a mask image appropriate for seeding streamlines on the gray matter-white matter interface. White matter tractography was used to estimate the fibre counts between each pair of voxels. The connection of a voxel to itself was set to 0. For BOLD signals, the following pre-processing was performed by fMRIPrep [14]: a skull-stripped reference volume was generated; a deformation field to correct for susceptibility distortions was estimated based on fMRIPrep's fieldmap-less approach; accordingly, a corrected EPI reference was calculated for a more accurate co-registration with the corresponding anatomical reference and then co-registered to the MNI standard space; the pre-processed data were smoothed with a full width at FWHM Gaussian kernel of 6 mm and filtered with band-pass filtering (0.01 ~ 0.1 Hz). For resting-state data, ICA-based automatic removal of motion artifacts (AROMA) was used to generate aggressive noise regressors as well as to create a variant of data that is non-aggressively denoised[15]. Finally, the pre-processed data were resampled at a resolution of 3*3*3 $mm^3$.

To integrate multimodal neuroimaging data into our neural network model more effectively, we applied a series of data cleaning procedures: (1) For cortical structures, including the cerebellum, only voxels with a gray matter volume (GMV) greater than 0.4 were retained. (2) For subcortical structures, including the brainstem, voxels with a GMV greater than 0.2 were kept. (3) Connections between voxels with a connectivity count of less than 2 were set to zero. (4) Isolated voxels, which do not form structural connections with any other voxel, were removed. Hence, a total of 23,863 voxels were included in the following model construction.

**Section 2: Details of constructing the Digital Brain**

*Details of the computational neuron model*

The computational neuron model of the Digital Brain could be generally a nonlinear operator from a set of input synaptic spike trains to an output axon spike train. In this work, the neurons were considered as the leaky integrate-and-fire (LIF) model [16]: A capacitance-voltage (CV) equation described the membrane potential of neuron $i$ when it is less than a given voltage threshold $V_{th,i}$.

$$C_i \dot{V}_i = -g_{L,i}(V_i - V_L) + \sum_u I_{syn,i} + I_{bg,i} + I_{ext,i},\ V_i < V_{th,i}$$

where $C_i$ was the capacitance of the neuron membrane, $g_{L,i}$ was the leakage conductance, $V_L$ was leakage voltage, $I_{bg,i}$ was the background current, and $I_{ext,i}$ was the external constant stimulus. It should be highlighted that $I_{ext,i}$ were the parameters, which were independently sampled following the Gamma distribution with certain hyperparameters to be assimilated. Four synapse types (i.e., AMPA, NMDA, $GABA_A$, and $GABA_B$) were considered in this model. The background current was given by independent Ornstein Uhlenbeck (OU) processes and described as follows:

$$\tau_{bg} dI_{bg,i} = (\mu_{bg} - I_{bg,i})dt + \sqrt{2\tau_{bg}}\sigma_{bg} dW_{i,t} \tag{1}$$

where $\tau_{bg}$, $\mu_{bg}$, and $\sigma_{bg}$ represented the uniform time-scale constant, the mean, and the standard variance of the Wiener processes $W_{i,t}$, respectively, which were uniform among neurons. When $V_i = V_{th,i}$ at $t = t_n^i$, the neuron registered a spike at time point $n$ and the membrane potential was reset at $V_{rest}$ during a refractory period. After that, $V_i$ is governed by the CV equation again. In the case that neuron $j$ was connected to neuron $i$, an exponential temporal convolution was established:

$$\begin{aligned} I_{u,i} &= g_{u,i}(V_u - V_i)J_{u,i} \\ \dot{J}_{u,i} &= -\frac{J_{u,i}}{\tau_i^u} + \sum_{k,j} w_{i,j}^u \delta(t - t_k^j) \end{aligned} \tag{2}$$

where $g_{u,i}$ was the conductance of synapse type $u$ of neuron $i$, $V_u$ was the voltage of synapse type $u$, $\tau_i^u$ was the time-scale constants of synapse type $u$ of neuron $i$, $w_{i,j}^n$ was the connection weights from neuron $j$ to $i$ of synapse type $u$, $\delta(\cdot)$ was the Dirac-delta function and $t_k^j$ was the time point of the $k$-th spike of neuron $j$.

*Cortical laminar model*

We considered the laminar model that was established based on the neuroanatomy of the primary visual cortex of the cat [17,18], which was widely used to build up large-scale neuromorphic computational models [19,20]. In this model, the cortex was composed of six layers: L1, L2/3, L4, L5, and L6, equipped with excitatory and inhibitory neurons.

At each layer, the number of neurons was proportional to the statistics given in Du. *et al* [19], which originated from Binzegger. *et al* [17]. Early studies with physiological experiments in area 17 of cats [21] and adult macaques [22] showed consistent results on the synaptic distribution of excitatory and inhibitory neurons in each layer [17]. We followed the data of assigned synapses in Binzegger. *et al* [17]. To treat the sources of the unassigned synapses, we made modifications to the data given the following assumptions as done in previous studies [17-20]: 1) all of the unassigned symmetric (i.e., inhibitory) synapses in layers 1-6 originate from smooth (inhibitory) neurons in the layer where the synapses were located (*2*); 2) 95% of the unsigned asymmetric (i.e., excitatory) synapses came from other cortical regions.

The numbers of the input synapses to each layer in the voxel coming from other layers and other cortical regions were proportional to the statistics given in Du. *et al* [19] as well. These actual numbers of input synapses of each layer could be calculated by fixing the average number of input synapses of neurons in this voxel as constant $d_{in}$. In the absence of sufficient biological information, we allocated the output synapses from the excitatory neurons across all layers based on their population sizes.

*Whole-brain structural network model based on structural MRI and PET data*

The network architecture of this computational model could be arbitrary including feedforward and recurrent. Herein, we proposed a general network model of the human brain, a hierarchical random graph with constraints and multiple edges (HRGCME), to represent the neuron pairwise synaptic connections, based on the biological neuroanatomy knowledge and structural neuroimaging data. Specifically, the HRGCME model is a set of weighted directed graphs with hierarchical structures:

$$\varsigma = \{G^{\rho} : \rho = 1,\ldots,P\}$$

where $G^{\rho} = (V^{\rho}, E^{\rho})$ with node set $V^{\rho}$ and edge set $E^{\rho}$, the corresponding adjacency matrix $W^{\rho} = \{W^{\rho}_{i,j} : i, j \in V^{\rho}\}$ and node attributes $A^{\rho} = \{A^{\rho}_{i} : i \in V^{\rho}\}$. We further defined a disjoint splitting of the node set $V^{\rho}$, $\{C^{\rho'}_{l} : l = 1,\ldots,|V^{\rho}|\}$, namely satisfying $C^{\rho'}_{li} \cap C^{\rho'}_{lj} = \emptyset,\ li \neq lj$ and $\cup_{l} C^{\rho'}_{l} = V^{\rho}$. Here, $\rho = 1$ represented the scale of neurons and a large $\rho$ could stand for the scale with lower resolution. There were two types of neurons: excitatory (pyramidal) neurons and inhibitory (inter) neurons. It should be noted that edges of $E^{u,1}$ with $u$ = AMPA, NMDA occurred only from excitatory neurons to other neurons and those with $u$ = $GABA_A$, $GABA_B$ occurred only from inhibitory ones. Self-loop and multiple connections for a pair of neurons were not allowed. Hence, $G^{u,1}$ shared the same node sets and edge sets for all $u$. The corresponding weights of the graph for different synaptic types were different but obeyed the same distribution (uniform distribution from 0 to 1 in our model). $\rho = 3$ corresponded to a network of voxels, where $V^3$ was the set of all brain voxels, $C_l^3$ was proportional to the gray matter volume of voxel $l$ measured by the VBM, and the element $C_{i,j}^{u,3}$ was consistent with values of the fibre counts between voxel pair $i, j$ obtained from DWI. The whole brain was further divided into four heterogeneous parts (i.e., cortex, subcortex, brainstem, and cerebellum). The ratio for each voxel between the number of excitatory synapses coming from this same voxel and the number of afferent synapses coming from the external voxels was set 2:1 in the non-cortical area, the same as cortical-column structure[23-25] The synaptic density for each voxel could be measured by the PET imaging of the synaptic vesicle glycoprotein 2A [26,27]. Hence mean ratios of inner-voxel over all connections were set as around 6/25, 14/25, and 16/125 for subcortex, brainstem, and cerebellum respectively. As for $\rho = 2$, each cortical voxel was regarded as a micro-column of six laminar layers with eight disjoint splitting populations (layer 2 and layer 3 were merged to L2/3 and excluded L1, each layer was split into 2 populations): $C_l^2$, $l = j_1, \ldots, j_8$. Each $C_l^2$ was proportional to $C_j^3 * T_l * D_l$, where $T_l$ was the thickness of layer $D_l$ was the neuronal density of population $l$. The connections between populations (long-range connections) were all excitatory, namely, $E^{u,2}$ and $W^{u,2}$ were empty when $u$ = $GABA_A$, $GABA_B$. The connection matrix of $G^2$ was derived by applying the Kronecker product of the normalized structural connectivity matrix (i.e., $G^3$) and micro-circuit matrix (Extended Data Table 3):

$$p(i \leftarrow j) = \{DWI \otimes M\}_{i,j}$$

where $p$ denoted the derived connection probability from population $j$ to population $i$. Altogether, we implemented this network model based on an extension of $k$-random graph [28]. Firstly, the number of neurons per population was assigned, Secondly, for each neuron, the number of synaptic inputs from each neuron population was calculated according to its property and location defined in $G^2$. Thirdly, each source neuron in the neuron population was sampled with equal probability without replacement iteratively.

The whole brain modelling in DB was divided into 4 parts: cortex, subcortex, cerebellum, and brainstem, and weighted 19%, 0.5%, 80%, and 0.5% respectively. Each voxel was embedded with a certain number of neurons, and that size was proportional to its corresponding gray matter information. Formally, we used $S_m$ to stand for the gray matter volume for brain voxels and $T_k$ represented the thickness of each layer. The fraction of excitatory neurons in each layer was taken to be identical across voxels at 80%. Hence, we can easily derive the size of neurons in each population $N(i)$ by the predefined network size $D$ as

$$N(i) = \begin{cases} \frac{S_m}{\sum S_j} \times \frac{T_i}{\sum T_i} \times D \times \Lambda_i \times (0.8\ or\ 0.2) & i \in C_m^{2,3}\ and\ |C_m^{2,3}| = 8 \\ \frac{S_m}{\sum S_j} \times D \times \Lambda_i \times (0.8\ or\ 0.2) & i \in C_m^{2,3}\ and\ |C_m^{2,3}| = 2 \end{cases} \quad (1)$$

Here, $|C_m^{2,3}| = 8$ implied a cortex voxel with a laminar model, and $|C_m^{2,3}| = 2$ implied a non-cortex voxel, where multiplied with 0.8 standing for excitatory neurons and 0.2 for inhibitory neurons.

The edges of the whole brain in $G^2$ were organized into two classes: local connections and cortico-cortical connections. The cortico-cortical connections were derived by applying the row-normalized matrix to distribute the outer connection of each voxel (DTI). Layer-specific tracing data from the inner connection data (denoted as $M_{i,j}$) helped us determine the distribution of connections across target layers and the source distribution. The local connection is derived from the row-normalized inner circuit data (Extended Data Table 3). Specifically, we write the formula as

$$p(i \leftarrow j) = \begin{cases} \dfrac{M_{i,j}}{\sum_k M_{i,k}} & if\ i,j \in C_m^{2,3} \\ \dfrac{M_{i,cc}}{\sum_k M_{i,k}} * \dfrac{DTI_{m1,m2}}{\sum_k DTI_{m1,k}} & if\ i \in C_{m1}^{2,3}, j \in C_{m2}^{2,3}, m_1 \neq m_2 \end{cases} \quad (5)$$

where $\mathrm{i}, \mathrm{j}$ represented the index of population respectively and $p$ denoted the derived connection probability. For combined local and long-range connections, we assumed the average number of incoming synapses $d_{in} = 1000$ for each voxel in the cortex and subcortex but $100$ in the brainstem and cerebellum. Then we assigned the average in-degree for each voxel according to the synapse density information. In detail, we calculated the in-degree for population $i$ as

$$d(i \leftarrow \cdot) = (100 \text{ or } 1000) \times \frac{\mathrm{syn_m}}{\sum_k syn_k} \times \frac{N_i}{\sum_k N_k} \times \frac{\sum_j M_{i,j}}{\sum_{k,l} M_{k,l}}, \quad where\ i,j \in C_m^{2,3} (6)$$

According to Eq. 5 and 6, the number of synapses of population i from population j for each neuron in $G^1$ was calculated as:

$$\mathrm{N_{syn}(i \leftarrow j) = d(i \leftarrow \cdot) \times p(i \leftarrow j)}$$

In summary, the basic setup data of the whole-brain neuronal network model were given in Extended DATA Table 1. It should be highlighted that we set different $\bar{d}$ from $10^2$ up to $10^3$, which is possibly different from brain regions, due to the limit of the communication and memory bandwidths. It has been widely known that generally the average synaptic number of each neuron is estimated from $10^3$ to $10^5$. Since the main simulation cost comes from communicating and reading spikes between neurons, it cannot be afforded a large synaptic degree for a whole-brain-scale neuronal network simulation by our GPU-based high-performance computer (HPC).

*Parameter initialization*

Parameter initialization for the Digital Brain's resting-state spike dynamics was achieved through an approximate off-network method (Supplementary Algorithm 1). This process ensured biologically sound spiking behaviours and the network's self-sustaining dynamics while accounting for heterogeneity. By extracting the smallest network units that characterized the infrastructure and employing an iterative approach, the conductance parameters of synapses were updated until the unit fired with a stable frequency. The method was semi-automatic and converged in less than 20 iterations.

The process involved modulating membrane potential through synaptic currents, allowing computation of excitatory and inhibitory currents based on desired firing rates. The conductance parameters were then adjusted to approximate the calculated current values. This approach ensured the large-scale network's output spike trains were stable at the specified frequency (10-20 Hz in this work).

We initialized parameters according to the spike dynamics of networks in the resting state (without specifically external inputs) numerically, according to the following aspects: 1) Distribution of the firing rates concerning neurons; 2) Response map of firing rate concerning the synaptic conductance parameters; 3) asynchronization between neuron spike trains. The first item was to guarantee biologically sound spiking dynamic behaviours, the second was to guarantee that the spike dynamical behaviours were mainly generated and maintained not by injected spike noises but by the network structure, and the third was to guarantee that it was not the initial homogeneous parameter but the heterogeneity of the network structure accounts for the neuronal spiking dynamics.

We utilized an approximate off-network method to initialize the parameters (i.e., the conductance parameters) to enable the network to simulate spike trains of a stable frequency and normal variance. Given the large-scale network configuration, we extracted the smallest unit that could characterize the network infrastructure. In this case, the entire network could be expressed as a complex coupling of such units (such as a single neuron in a sparse network or, a micro-column in a cortical laminar network). Considering a small network with some such units, our goal was to find a stable parameter so that the unit could fire with the same frequency in the network under a given spike input. We injected statistically independent Poissonian spike trains with a fixed frequency (5-10 Hz) to each unit and adopted a semi-automatic iterative method to update the synaptic conductance parameters until all selected neurons stably fire with the same frequency (10-20 Hz) or up to a limit iteration (we set 20). From the iteration process, we could see that the off-line network gradually approaches the firing state specified by us, in which parameters converged in about ten iterations. At last, the derived parameters were set in the large-scale network simulation. It has been seen by the network simulation that the output spike trains are approximately stable at the given frequency.

In detail, for a given large-scale network model, we aimed to find a set of proper parameters (explicitly, conductance parameters of synapses $g_{ui}$) for the spiking network to maintain a stable firing rate. To do this, we reduced the network size while maintaining consistent topology and neuron properties, and adopted the semi-automatic iterative method in the off-line network to update the parameters, until the network converged.

Reviewing the CV equations (Eq. 1), the membrane potential was modulated by 4 synaptic currents and the leaky current:

$$\mathrm{C}\dot{V} = I_{leaky} + I_{AMPA} + I_{NMPA} + I_{GABAa} + I_{GABAb}$$

where $\mathrm{I}_{\mathrm{leaky}} = -g_{leaky}(V - V_{leaky})$ depended on the membrane potential and was approximated as $\mathrm{I}_{\mathrm{leaky}} = -g_{leaky}(\frac{V_{th}+V_{rest}}{2} - V_{leaky})$ in the evolution process. So, given the allowed maximum fire rate of the network, we calculated the EPSP $\mathrm{I}_{\mathrm{exc}}$ for the consideration that the network fires at a maximum level if the inhibitory synaptic current is zero. Then we calculated the IPSP $\mathrm{I}_{\mathrm{inhi}}$ while the network fires at a stable firing rate $\mathrm{r}_{\mathrm{normal}}$. Empirically, to prevent the network from collectively oscillating, we set the ratio of the 2 excitatory synaptic currents to be 0.3 and 0.7 respectively (reducing the ratio of excitatory current of fast channels). The ratio of the two inhibitory channels is set to 0.5 and 0.5. Now, we could compute the synaptic current for 4 chemical channels respectively according to the EPSP and IPSP. Empirically, the synaptic currents were approximately linear about synaptic conductance, so we could tune parameter $\mathrm{g}_{\mathrm{ui}}$ to enable the synaptic current close to the PSPs value that we calculated above.

**Section 3: Parallel operations of computation and communication in Digital Brain simulation**

*Hardware and software environments.*

The C++ program using C++ compiler (g++, version 7.3.1) was developed with the MPI library functions and rocm-4.1.0 hipcc compiler for GPU. In addition, the Python program for HMDA was developed using Python version 3.7.11. The Python and C program runs in a Linux environment (CentOS 7.6.1810). The simulation is performed on 3503 computing nodes in the cluster in Advanced Computing East China Sub-Centre. Each node has a single 32-core processor operating at 2 GHz and 128 GB of DRAM memory and 4 GPUs operating at 1.319 GHz with each GPU having 16GB HBM2 working on 800MHz, as well as 819.2GB/s memory bandwidth. The communication between GPUs within one node is through shared memory, while communication between nodes is through a 200Gbps Full Duplex Infiniband network. Thus, the total theoretical peak computational capability is 151.3296 PFLOPS, and the total amount of GPU memory is 224.192TB.

*Numerical method.*

For resource and computational efficiency in modelling, the numerical computation scheme employed in this study comprises two components: the first-order Euler-Maruyama time-stepping method and the delayed spike recording mechanism, despite another numerical method of spike integration with better performance[29].

Specifically, we distinguish between the time-step for iteration, denoted as $\Delta_t$, and the time-bin for transferring spikes, denoted as $b_t$. The detailed algorithm is formulated as follows.

$$\mathrm{J_{u,i}}(t+\Delta_t) = \exp\left(-\frac{\Delta t}{\tau_i^u}\right) J_{u,i}(t) + \sum_j w_{i,j}^u A_j(t), \quad (3)$$

$$I_{bg}(t+\Delta_t) = I_{bg}(t) + \frac{\mu_{bg} - I_{bg}(t)}{\tau_{bg}}\Delta_t + \sqrt{2\tau_{bg}\sigma_{bg}\epsilon_t}, \quad (4)$$

$$V_i(t+\Delta t) = V_i(t) - g_{L,i}(V_i(t) - V_L)\Delta_t + \sum_u J_{u,i}(t) g_{u,i}\big(V_u - V_i(t)\big)\Delta_t + I_{bg}(t), \quad (5)$$

$$V_i(t+\Delta t) = V_{rsy}, \quad when\; t+\Delta_t \in \left[t_i^s, t_i^s + T_{ref}\right].$$

where Aj (t) represented the accumulated spike of the neuron j:

$$\mathrm{A_j}(t) = \begin{cases} 1, & if\; t = k \cdot b_t\; and\; j\; neuron\; emits\; spike\; in\; [t - b_t, t] \\ 0, & else \end{cases}$$

This spike recording approach minimizes communication overhead and significantly enhances numerical computing speed.

Thus, we set $b_t = 1$ $ms$ in all cases (even when $\Delta_t = 0.1$ $ms$). We have validated that the error of this canonical Euler-Maruyama algorithm is ignorable tiny in both membrane potential and spiking time during the period 800 ms, i.e., the temporal resolution of the BOLD signals in the experiment, as shown in Extended Data Figure 1. Hence, it contributed no further biases when conducting data assimilation to the fMRI data. Given that the spikes of each neuron could be rather sparse and irregular, we provided a destination-address list for each neuron as the complement of the spike-driven scheme (Supplementary Algorithm 2). Although it cost additional time on atomic operations, it was still much faster than the natural pull-based algorithm in practice because it took full advantage of both the sparsity of $A_j(t)$ and $w_{i,j}^u$. To initialize the spike trains and avoid synchronization of neuronal spikes, we initially set the homogeneous values for each type of parameter and then injected a background spike into the output spike train of each neuron with a low independent identical probability from 0.5% to 1%, turning out an external spike frequency from 10 to 20 Hz.

*Data structure*

Representing the neural network is a crucial task to address before deploying the model. Here, we leverage the sparsity of network connections, utilizing sparse graph representation and relative numbering methods to depict the brain graph network, significantly reducing the storage space required for network representation. Specifically, the representation of neurons on each card is divided into two tables (as shown in Supplementary Figure 8 and Supplementary Table 5): one containing all parameter attributes for all neurons on that card (properties), and the other containing all incoming connections to neurons on that card as target neurons (connection table). Regarding the connection table, the index of the target neuron is decomposed into the card number and the relative neuron number to reduce the integer representation limit of neuron indices. Specifically, for each connection edge, the GPU card number, relative neuron index, excitatory or inhibitory type of the edge, and weight parameters of the edge are recorded. Overall, this storage and representation is summarized in Supplementary Table 5 and significantly optimizes storage space and memory usage.

*Parallel operations of computation and communication.*

GPU hardware advancements have significantly enhanced performance across various domains, including image processing, deep learning, and large-scale data processing. Furthermore, they have found applications in simulating spiking neural networks. In this project, we utilize GPUs to implement a distributed simulation of spiking neural networks, specifically the DB model. Applications harnessing GPU devices adhere to a paradigm where operations to be performed on data are encapsulated as "kernels". These kernels, when executed, process individual data elements concurrently across numerous processing threads on the GPU. Within the context of DB simulation, these kernels are tailored to execute tasks such as updating neuron membrane voltages and translating neuron spikes into relevant synaptic inputs.

Eq. (3) has been decomposed into the following three parts: the second term in the original equation distinguishes between inter-card spike calculation and intra-card spike calculation, which are then summed with the variable exhibiting exponential decay in the first term.

$$\mathrm{J_{u,i}}(t+\Delta_t) = \exp\left(-\frac{\Delta t}{\tau_i^u}\right) J_{u,i}(t) + \sum_{j \in GPU(i)} w_{i,j}^u A_j(t) + \sum_{j \in else\ GPU} w_{i,j}^u A_j(t) \quad (6)$$

The purpose of the decomposition in the above equation can be understood from two perspectives. Firstly, from a computational standpoint, spikes belonging to the same GPU card inevitably arrive at downstream neurons much faster than others. The separation of these spikes facilitates separate computations, thereby optimizing thread operations rather than waiting for external pulses to be fully received before initiating calculations. Secondly, from a physical perspective, intra-card communication is achieved through shared memory, while inter-card communication involves complex copying and communication via the "GPU-CPU-CPU-GPU" chain route, necessitating the separation of the kernels for both calculations.

We have achieved maximum parallel computation by separating the computational and communication threads. Below, we will provide a detailed introduction to these kernel definitions and the operations they implement.

**The computational kernel** is responsible for managing the numerical computation of neural membrane equations on the GPU within the card. Specifically, the GPU first reads the state variables from memory to update the neuron's Ornstein-Uhlenbeck (OU) current (as in Eq. (4)), followed by updating the membrane potential and determining whether a spike is fired (as in Eq. (5)). The Null stream handles the computation, while simultaneously activating two additional streams: one invokes the GPU's Memory Control to copy spike information to CPU memory, and the other is responsible for accumulating and tallying spikes transmitted from active synapses within this card (as in Eq. (6)). Upon completing the aforementioned tasks, the stream pauses until it receives spike sequences from other GPUs, then proceeds to complete the remaining operations in Eq. (6), thereby finishing one complete numerical iteration.

**The sending kernel** is responsible for sending spike information to other CPUs. Once the spike sequence is copied from the GPU to CPU memory, the spikes of pre-synapse are divided into multiple packages and sent to the corresponding downstream neurons' CPUs. Inter-CPU communication here utilizes MPI (https://www.open-mpi.org/), employing custom secondary routing to facilitate point-to-point communication. Routing strategies ensure efficient utilization of IB bandwidth while forwarding aggregation helps mitigate communication contention and congestion to some extent. Given the need to calculate the average firing rate of different neuron groups and even collect status variables such as neuron membrane potentials and synaptic currents in network model simulations, nodes also need to send information to the main node (host node). The transmission speed depends on the amount of data being sent and whether the main node is experiencing communication congestion.

**The receiving kernel** is responsible for receiving information from other CPUs, storing it in CPU global memory, and copying it to the GPU. As some nodes with lighter workloads may quickly complete their computations and communicate outward, the receiving kernel of each node is launched at the beginning of each iteration to avoid information loss. Once all information from other cards has been fully received and copied to the CPU, the computational thread continues to initiate subsequent calculations.

**Section 4. Voxel-GPU mapping and two-level routing**

To achieve the data traffic management problem, we presented an optimization framework to improve the delivery of hyper-giant traffic in brain simulations, focusing specifically on the mapping between voxels and GPUs. We employed a low-latency communication design method to optimize the efficiency to simulate the whole-brain neuronal model on GPUs with Open-MPI [30]. We focused on the flow of logical data in an HPC composed of GPU nodes. Since the physical network architecture cannot be changed, when designing voxel-GPU mapping for a system, we considered some limitations including hardware conditions (such as bandwidth limitations) and frameworks on HPCs.

Both the partition algorithm and routing method need to consider balancing the data traffic among GPUs (Figure 1c). Owing to the limitations of hardware and the different connected relationships between neurons, planning the partitioning of neurons into different GPUs and designing the routing of the communication between these GPUs could increase hardware utilization and connect more of the information between neurons in the same GPU. Furthermore, the number of connections across different GPUs was reduced to minimize the connections of the entire network. An increased number of connections across different GPUs could result in memory overhead in the network, causing time delay and affecting even the normal operation of the system. When the number of connections was smaller, the system would incur fewer overheads, while performing the task.

Using the partitioning algorithm, we may assign neurons to GPUs in the system based on the amount of data with which they interact. However, owing to the difference in the amount of data information contained by the neurons, differences still existed in the amount of data between different GPUs. Without proper traffic processing, thousands of GPUs in an HPC that were used to simulate the human brain would communicate with each other simultaneously, which inevitably led to network congestion. The GPUs all needed to communicate with a GPU at the same moment (Figure 1c), if the amount of data generated by one GPU in the simulation process was more than that generated by other GPUs for the total system running time, the result should only be obtained after all GPUs have fully executed. Hence, when the system was running, the resources of the majority of GPUs were wasted, which was not conducive to minimizing the running time of the entire system. Thus, balancing the amount of data generated in each GPU as far as possible could promote the improvement of the low-latency communication design in brain simulations. Furthermore, as the simulation was run on the HPC, communications among a large group of GPUs only adopted the functions in the MPI architecture; therefore, to control the data traffic, we further propose the two-level method to perform the logical hierarchical processing between the GPUs.

*Partition algorithm*

As mentioned above, the proposed partitioning scheme assigns neurons to GPUs to minimize the traffic between different GPUs. Since communications between neurons on the same GPU are much faster than that between neurons across GPUs, this algorithm reduces communication latency by assigning a set of neurons that have strong communication demand between each other on the same GPU. Furthermore, this algorithm makes the traffic between each pair of GPUs in the system as balanced as possible to avoid network congestion thus improving delay performance.

The objective of neuron partition in brain simulations is to assign neurons to GPUs to balance the traffic load for inter-GPU communications; that is, minimize the traffic volume of the inter-GPU communication session with the heaviest load. Therefore, mapping $n$ neurons to $N$ GPUs in a brain simulation can be formulated as an $N$-way partition for a graph; that is, finding a partition $P = \{V_1, V_2, \cdots, V_N\}$ that assigns n neurons to $N$ GPUs, to minimize the traffic volume on the edge between the pair of GPUs that has the heaviest traffic load. The huge number of neurons with their interconnections in the whole human brain simulation presents a scalability challenge to this partitioning problem. In neuron-level human brain simulations, the n interconnected neurons can be modelled as a weighted directed graph $G = (V, E)$, where the neurons are represented as vertices $V = \{v_1, v_2, \cdots, v_n\}$ and the connection between a pair of neurons $v_i$ and $v_j$ is represented by the edge $e_{i,j} \in E$, $E \subset V \times V$. The weight of vertex $v \in V$ is the size of the corresponding neuron and the weight of edge $e_{i,j}$ is the interconnection probability for the pair of neurons $v_i$ and $v_j$.

Therefore, we can formulate the neuron partition problem to minimize the following function, where $D_{ij}$ denotes communication traffic from $V_i$ to $V_j$:

$$F(P) = \max_{j} \sum_{i=1}^{N} D_{ij}$$

The partition solution $P$ to the neuron partition problem in brain simulations is subject to some constraints. First, the solution $P$ must divide $V$ into $N$ disjoint subsets:

$$\bigcup_{i=1}^{N} V_i = V, V_i \cap V_j = \emptyset \; \forall i \neq j$$

Secondly, the limited capacity of each GPU requires that the sum of the neurons' sizes simulated by each GPU is lower than a predefined upper bound $\gamma$.

While maintaining the balance of the volume of each GPU, we use a greedy strategy to assign neurons to every GPU to ensure high cohesion among neurons within a GPU and low coupling between neurons on different GPUs. It is worth noting that this greedy assignment is achieved by optimizing traffic between neurons (which is denoted as $D$), and the detailed calculation method of this traffic can be referred to in paper[31]. Finally, the algorithm updates the solution and outputs a table for neuron-GPU mapping.

To demonstrate the effectiveness of partition algorithms on human-scale networks, a comparison of inter-GPU synapses simulated with different neuron partitioning methods is presented in Figure 1d, where the red and green columns give the histograms of frequency distribution for the total number of inter-GPU synapses generated in the simulation system using the sequential method and the proposed partition method, respectively. It can be seen from the red column that the sequential method causes significant fluctuation in the amounts of synapses from different GPUs, which may easily cause traffic congestion on some network links in the supercomputer. The green column shows that the proposed partitioning algorithm may greatly reduce the difference in synapse numbers across GPUs. Therefore, the proposed algorithm is more effective in balancing traffic and avoiding congestion compared to the sequential method.

*Two-level routing method*

As mentioned above, the two-level routing method proposed in this paper was used to balance the communication speed and degree of congestion between GPUs. The routing for data traffic in brain simulation here was set to two levels; that is, communication between any two GPUs could only be forwarded at most once. Such a structural design satisfied all the restrictions mentioned above as much as possible while balancing the number of connections and forwarding time between GPUs and the traffic between all GPUs in the system as much as possible. The GPUs in the system were divided into groups according to the amount of data that should be communicated between GPUs. The amount of data that needed to be exchanged between the GPUs in the same group was relatively large, and the amount of data exchanged between the GPUs in different groups should be as little as possible. GPUs in the same group communicate with each other through direct connections, and GPUs in different groups need to identify the corresponding GPU node in their group to forward data traffic.

While a GPU is connected to a GPU in another group, the GPU needs to judge whether it can communicate with it on behalf of its group. If the answer is no, it needs to find another GPU as the bridge node to transmit its information. The amount of data that needed to be exchanged between the GPUs in the same group is relatively large, and the amount of data exchanged between the GPUs in different groups should be as little as possible. The reason was that GPUs in the same group communicated with each other through direct connections, and GPUs in different groups needed to identify the corresponding GPU node in their group to forward data traffic. Moreover, a group of GPUs on the same switch in the physical structure speeded the information interaction. In addition, because each connection required a thread to be started, the time taken to start the thread in the whole system could be reduced by reducing the number of connections. By matrix reordering, GPUs with dense connections or no connections could be grouped to reduce the number of connections (Figure 1d).

**Section 5. Further results of the performance of Digital Brain simulation**

We performed a spiking neuronal network simulation of the human-brain scale, using 3503 computer nodes and 14012 GPUs. The whole brain model consists of 86 billion neurons and 47.8 trillion synapses (AMPA, NMDA GABAa, and GABAb) with the topological structure mentioned in the neuron model section. In this model, the number of neurons loaded on each GPU is dominated by the GPU memory size. Due to the simulator storing enough neuron properties as configurable parameters to meet custom patterns, the data structure determines that up to 61.3 million neurons are loaded on each GPU in this subsection.

Actually, according to the synchronous scheme, the total simulation time cost depends on the GPU of the slowest computation thread. Thus, we collect all-time statistics during 4s simulation by statistical interface and define total simulation time using the last 800 msec (the period of the functional MRI data in our experiment) data.

*Statistical interface*

Timings of computation and communication are done manually by recording time anchors. We have designed a performance statistics interface to collect time anchors and Flops of each GPU, as shown in Supplementary Table 1. The parallel operations of computation and communication over the three major threads, as described in the main text, cause the overlaps between these timing that decrease the total simulation duration, as shown in Supplementary Figure 3.

The entire calculation cycle starts with equations (1,2) update at $t_0 = 0$ and ends with completing summing the synaptic currents $I_{sum}$ update at $t_3$. Thus, the total simulation time is recorded as $T_{sim} = t_3$. The computation time cost is named $T_{com} = t_3 - t_2 + t_1$ and the sending time cost $T_{send} = t_9 - t_8$. Because the time anchor $t_1$ for different GPUs to complete the membrane potential update and spike information copied is different, the start time of sending threads of different GPUs is also different. The receiving thread needs to start as early as possible, so it is sending cost time rather than receiving time cost $T_{rec} = t_{11} - t_{10}$. Furthermore, we calculate the intra-node communication time $T_{\cdot,intra}$ and inter-node communication time $T_{\cdot,inter}$ of spike information and it is obvious that $T_{\cdot,intra} + T_{\cdot,inter} = T_{send}$.

The real-time factor is defined as $T_{tos} = T_{sim}$ for each second in the equations (1,2) named by the biological time. For instance, in the case of the whole brain network with an average firing rate of 7Hz, the simulation time for 1 sec of biological time is 65s, which means $T_{tos} = 65$.

*Week scaling experiments*

An important advantage of our simulator is to achieve a large number of parallel calculations to update neuron equations which can be distributed to a large number of GPUs. Weak scaling relates to scaling the network model in such a way that the number of neurons loaded on each GPU remains relatively fixed as the number of nodes is increased. Perfect weak scaling yields that the computation time remains unchanged across any network size. In this subsection, the number of neurons on each GPU is fixed at 15 million. By varying the number of GPUs from 20 to 1000, different scales of cerebral cortex models with the same network topology have been simulated at an average firing rate of 7.6 Hz, as shown in Supplementary Figure 4.

Due to the increase in the number of GPUs, the same spike information needs to be sent to different GPUs, resulting in an increase in duplicate parts during spike information copied. This in turn leads to an increase in total communication traffic, increasing the total communication time cost. Besides, according to Supplementary Figure 4, computation time seems to increase as well. We believe that is because the computation time on each GPU is different, and the computation time cost depends on the slowest GPU. In the ideal case of weak scaling, computation time cost remains unchanged, and the time to solution slightly increases due to the increase in communication time, also as shown in Supplementary Figure 4.

*Strong scaling experiments*

It is believed that the computational performance of supercomputers has been increasing exponentially. Thus, strong scaling is to study the effect on simulation time for a fixed brain scale with increasing computing power, which is represented by the number of GPUs. Perfect strong scaling yields that the computational time halves by doubling the computational power. For the fixed scale of the cerebral cortex model with 1.5e6 neurons, the number of GPUs is changed from 100 to 1000, as shown in Supplementary Figure 5. As the number of GPUs increases, the computation time cost continues to decrease, approximate inverse proportional function.

*Different firing rates*

As mentioned above, communication traffic transferred between nodes mainly consists of spike information. So, the increase in the firing rates of the brain network leads to an increase in spikes, further leading to an increase in communication traffic. Meanwhile, the calculation of synaptic currents sums up the spikes of presynaptic neurons. Theoretically, as an increase in firing rate, the total computation time gradually decreases in a linear relationship. Here we vary the firing rate of the network from 10 to 160 by adjusting $g_{AMPA}$ in the equations (1,2) in the main text. We have shown the relationship between the total simulation time and different average firing rates as shown in Supplementary Figure 6. It is evident that the higher the firing rate, the higher the total simulation time. When the firing rate is up to 169.9Hz, the total simulation time can reach up to 280 sec.

*Different synaptic in-degrees*

Here we assess the impact of varying input synaptic numbers per neuron of cortical and subcortical models on computational efficiency. To facilitate model scaling and variable control, unless otherwise specified, the large-scale network used in the rest of this section only models the cortex and subcortical regions as part of the whole brain.

To save computing resources, we built networks of smaller neuron numbers, i.e., 1.5 billion neurons with varying input synapse numbers per neuron from 100 to 1000. Because the storage of synaptic information also leads to an increase in memory, thus the GPUs used to implement networks also range from 100 to 1000. On one hand, an intuitive impact is an increase in the number of input synapses per neuron leads to an increase in computing resources, and the increment of the number of GPUs also leads to an increment of communication traffic as well as the total communication time. Supplementary Figure 7 also verifies. On the other hand, the increase in the number of input synapses per neuron implies that a single neuron would receive more spikes from presynaptic neurons. So, the floating-point operations of calculating the weighted sum of spikes also increase, resulting in longer computation time. According to the trend of the total simulation time curve in Supplementary Figure 7, if low computational efficiency can be tolerated, the average degree can reach up to 5000.

*Comparison with a CPU-based tool*

We conduct a comparison of the computational performances between our Digital Brain and the well-used tool: NEST (version 3.7.0) [34]. We employed the data in S1 and the method in S2 to generate spiking neuronal networks of diverse sizes of 2, 3, 4, 5, and 10 million neurons respectively, with an average synaptic degree of around 100. We used the initial parameters by tuning the AMPA synaptic conductance parameter to set the average firing rates as around 3 HZ and 10 HZ respectively. We implement the Euler-Maruyama method with time step 1 ms. As a comparison, the NEST is run on one node (with one CPU with 32 cores of 2 GHZ and 1.024 TFLOPS) in double-float and our digital brain is run on one GPU card (with 256 cores of 1.319 GHZ and 10.8 TFLOPS) in both single- and double-float. The hard- and soft-ware environment is the same as mentioned in S3. Each network model is run for 3 secs and the last second is to calculate the real-time factor for comparison.

We should highlight that due to the different network generate schemes between our Digital Brain and NEST, the network architectures are slightly different. Our model has the average synaptic degree as 100 but the NEST as 98, which caused differences in the mean firing rates even with the same parameters.

As shown in Supplementary Table 4, the GPU-based Digital Brain is around 10 times as fast as the CPU-based NEST in both the mean firing rates around 3HZ (2.7 HZ in the NEST and 4.1 HZ in our Digital Brain) and around 10 HZ (9.4 HZ in the NEST and 12.1 HZ in our Digital Brain), except the network of 2M neurons with around mean firing rate around 10 HZ. This quite coincides with the numbers of the peak FLOPS of the CPU and GPU that we were using. We also generate a network of 10 million neurons which cannot be run on the CPU by NEST but can be run on the GPU by our Digital Brain. Even though the CPU is faster than the GPU (2 GHZ versus 1.319 GHZ), our Digital Brain can fully make use of the higher peak FLOPS of the GPU to achieve much better performance when simulating spiking networks of "smaller" sizes than CPU-based NEST tool.

Furthermore, it's noteworthy that due to the limited neuron count supported by the CPU-based simulators, our comparisons were conducted on a relatively smaller scale. For instance, the CPU-based NEST cannot afford to simulate a network of 10 million neurons but our Digital Brain can. NEST has simulation times that are around ten times slower compared to our system using GPUs. However, simulation time is the most crucial metric in brain simulation, and shorter simulation times are beneficial for studying the impact of brain structure on brain function. Hence, we show that GPUs are a better choice than CPUs to establish a simulating system of very large-scale spiking neuronal networks.

**Section 6. Details for Hierarchical mesoscopic data assimilation (HMDA)**

The data simulation aimed to estimate the hyperparameters of all voxels by tracking the BOLD signals. Therefore, the general evolution equations (i.e., the framework of HMDA) could be established:

$$\begin{cases} x(t) = G\big(x(t-1), \theta(t-1)\big) + \varepsilon_g \\ z(t) \leftarrow SpN\big(x(t)\big) \\ r(t) = BW\big(z(t-1), r(t-1), \theta(t-1)\big) \\ y(t) = Hr(t) \\ \theta(t) = cdf^{-1}\Big(cdf\big(\theta(t-1), \vartheta(t-1)\big), \vartheta(t)\Big) \\ \vartheta(t) = \vartheta(t-1) + \varepsilon_\vartheta \end{cases}$$

where *x(t)* represents the state variables (such as membrane potentials and synaptic currents) of all neurons; *G(·)* gives the discrete-time integrals of CV equation; $\varepsilon_g$ represents the white Gaussian noises; *SpN(·)* stands for the threshold model of registering spike trains from the membrane potentials; *z(t)* gives the neural activity measured by spike counts; *BW(·)* is the Euler-Maruyama discrete-time version of the Balloon-Windkessel model with *r(t)* for its state variables; *y(t)* is the BOLD signal generated by the hemodynamic response function; *θ(t)* represents all parameters in the models to be estimated and *ϑ(t)* is the corresponding hyperparameters; *cdf(·)* is the cumulate distribution function of the parameters (alternatively the empirical distribution if taking them by bootstrap method, for example, Monte-Carlo Markov Chain (MCMC)). In practice, the parameter inference was realized by the ensemble

Kalman filter (EnKF), and the hyperparameter inference was realized by a simple random-walk bootstrap filter.

Thus, we established a framework for HMDA of fMRI over the neuronal network and hemodynamic models by taking all states of neuronal including neural activities (spikes), synaptic currents, variables of hemodynamics, bold signals, and the parameters to be estimated as well as its preassigned distribution with the hyperparameters (Supplementary Figure 1). Analysis and prediction of the HMDA filter were executed at each time point of the BOLD signals. The time scale follows the biological clock (in ms) and takes each time step as the period of the fMRI scanning (i.e., 800 ms). We utilized the diffusion ensemble Kalman filter for the population of state and parameter inference, which gave the framework of distributed hierarchical mesoscopic data assimilation (dHMDA). In the following work, we took the hyperparameters of the synaptic conductance $g_{u,i}$ in Eq. 3 to statistically infer the model with the resting-state fMRI data and those of the external current $I_{ext,i}$ in Eq. 1 to statistically infer the model with task fMRI data.

*Correlation coefficients*

We mainly used the Pearson correlation coefficients between the BOLD time courses of the simulation DB after data assimilation and the experimental fMRI data in voxels to measure the performance of the DB. Since the ensemble Kalman filter tracked the real data usually with small delays, we alternatively employed the following lag correlation coefficients:

$$pcc(x_{DTB}(t + lag), x_{exp}(t))$$

where $pcc(\cdot,\cdot)$ was the Pearson correlation coefficients, $x_{DTB,exp}$ represented the time course acquired by the DB and experimental counterpart, and $lag$ was the time delay for assimilation. We take $lag = 2$ throughout the paper, which is selected to maximize the average correlation in the resting-state experiment and fixed for the task experiment.

## Supplementary Figures and Tables

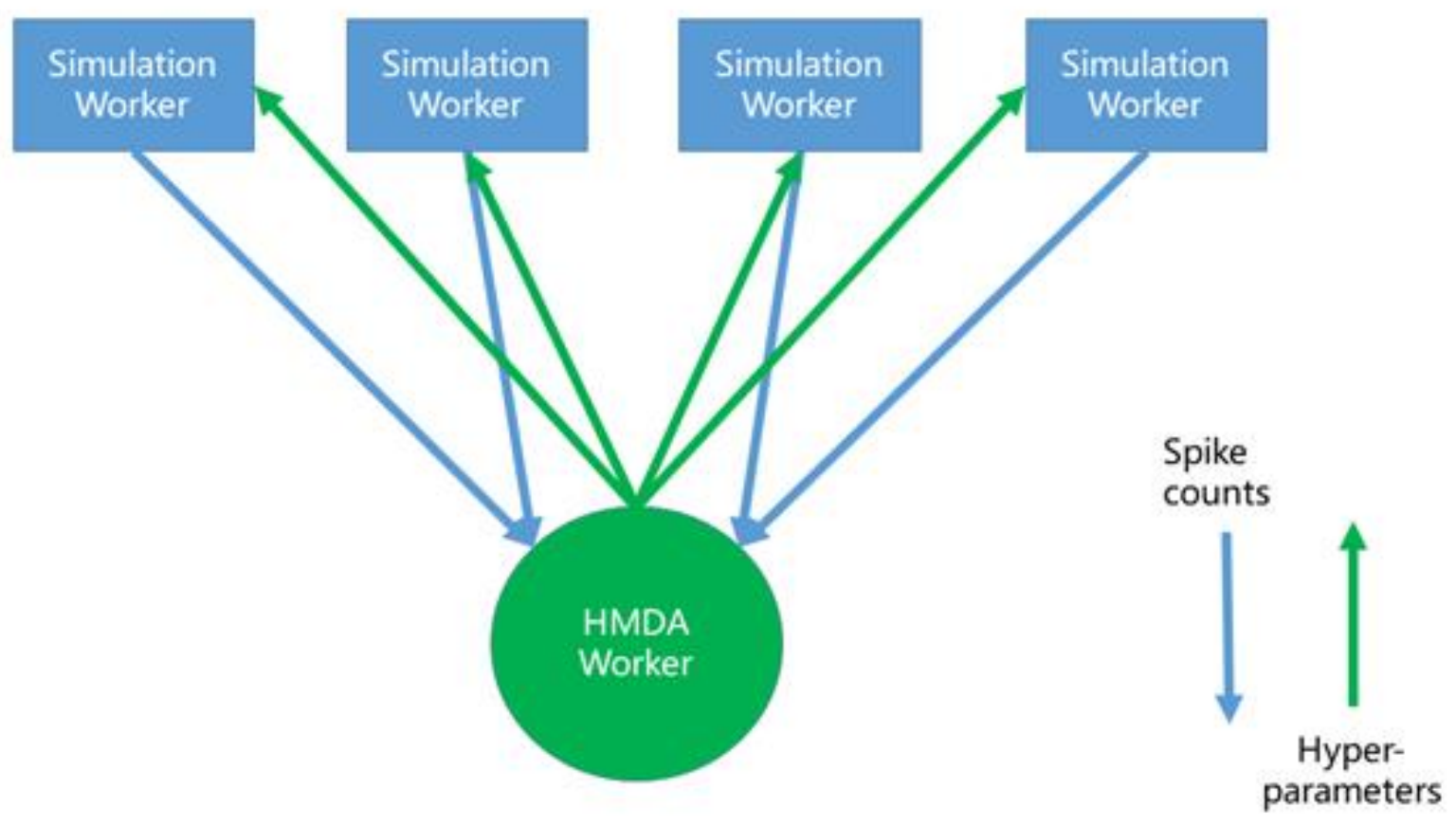

**Supplementary Figure 1. Layout framework of HMDA for fMRI data in an HPC system.** The central HMDA node updates the hyperparameters via the spike counts from all simulation nodes and distributes the updated hyperparameters to all simulation nodes (see Ref. 20 for details).

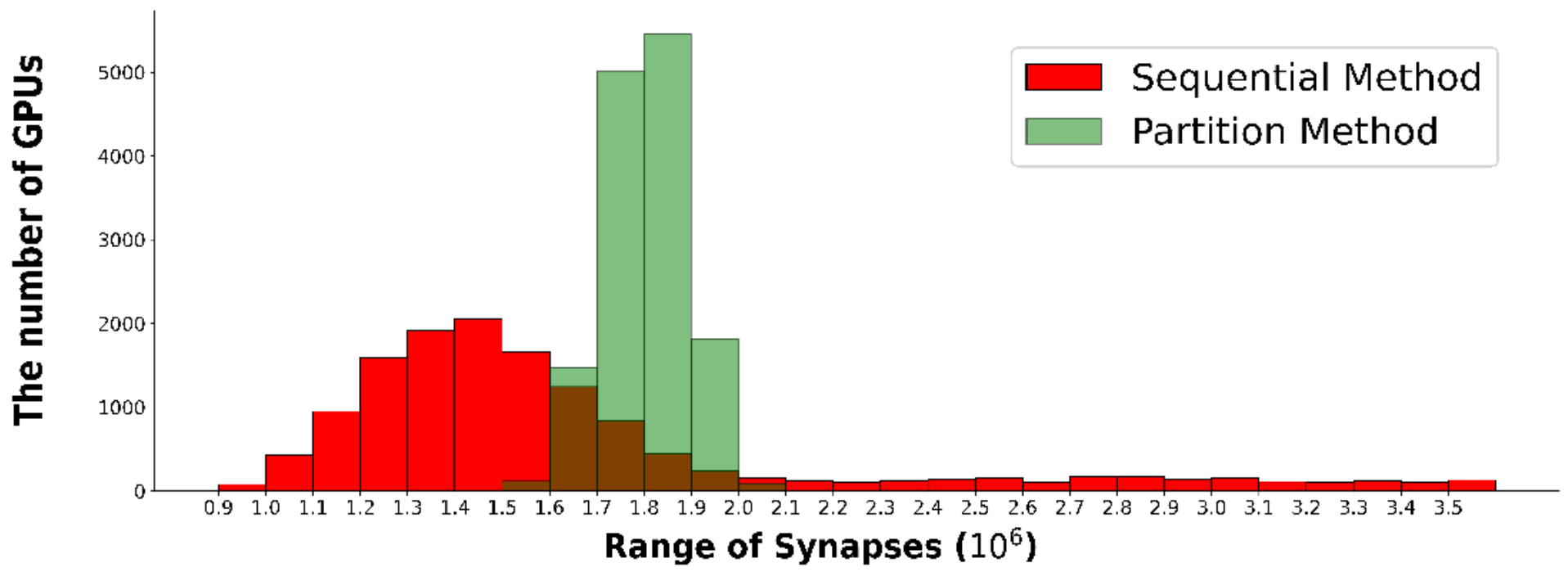

**Supplementary Figure 2. Comparison of inter-GPU synapses simulated with different neuron partitioning methods.** The red and green columns stand for the histograms of frequency distributions for the total number of inter-GPU synapses generated in the simulation system using the sequential method and the proposed partition method, respectively.

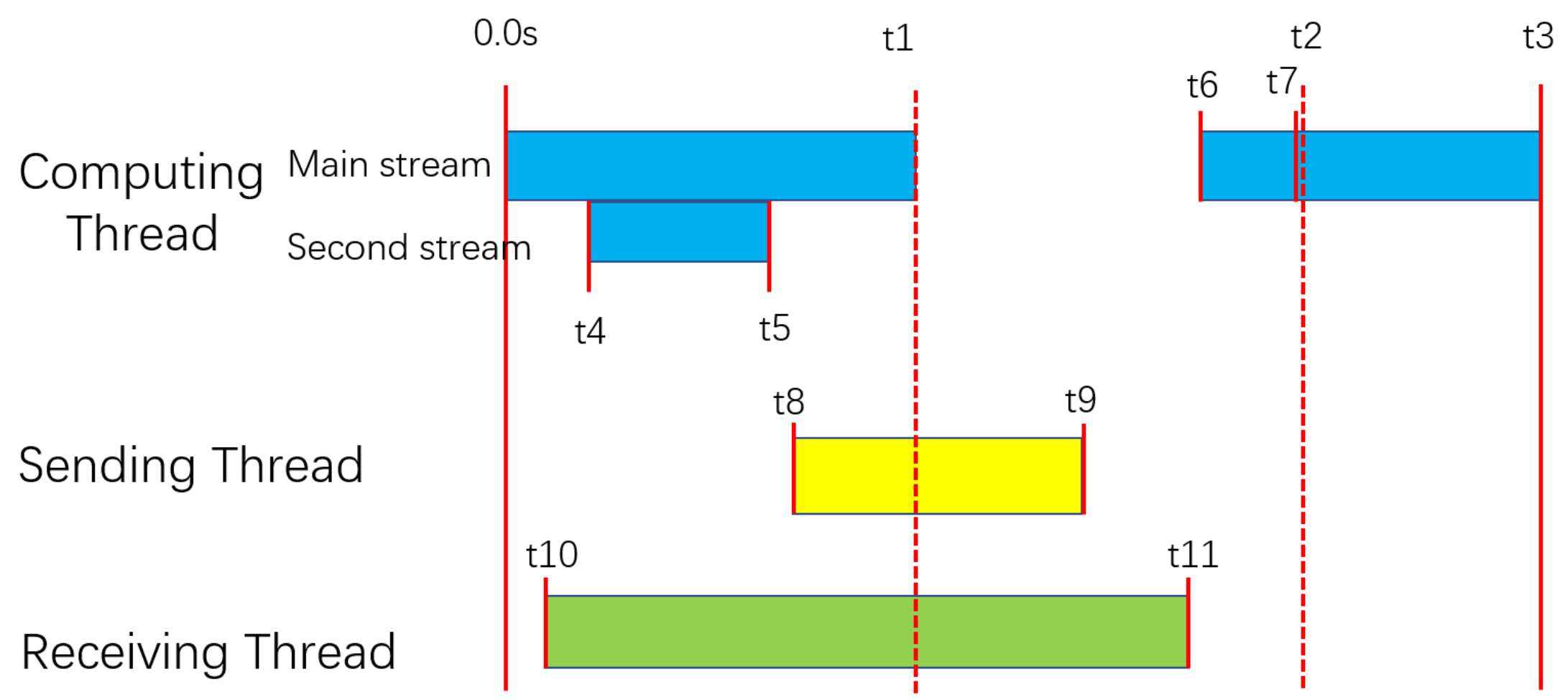

**Supplementary Figure 3. Schematic diagram of parallel operations and time anchors.** The overlapping between the computing, sending, and receiving threads in time illustrates the parallelization performance of our codes that implement the DB simulation. See Supplementary Table 1 for the notations $\boldsymbol{t_{1-11}}$.

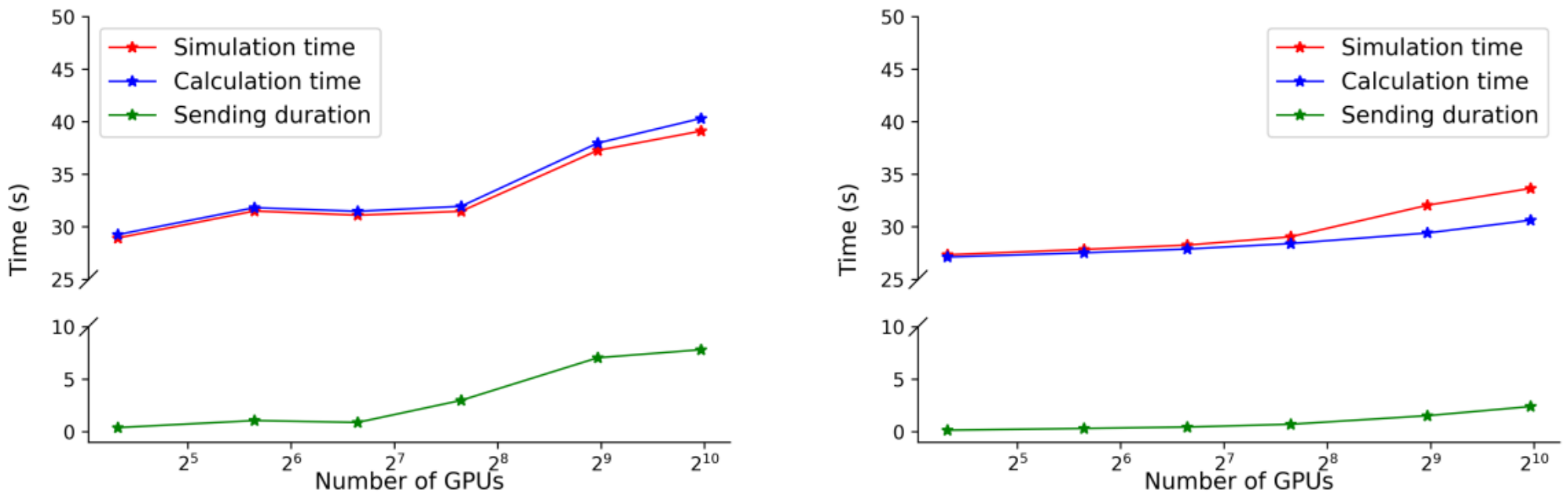

**Supplementary Figure 4. Weak scaling property**. The maximum (left) and average (right) time costs over all GPUs of simulation, computation, and sending in simulating one biological second are plotted with varying numbers of GPUs: 20, 50, 100, 200, 500, and 1000, with each GPU containing 15 million neurons.

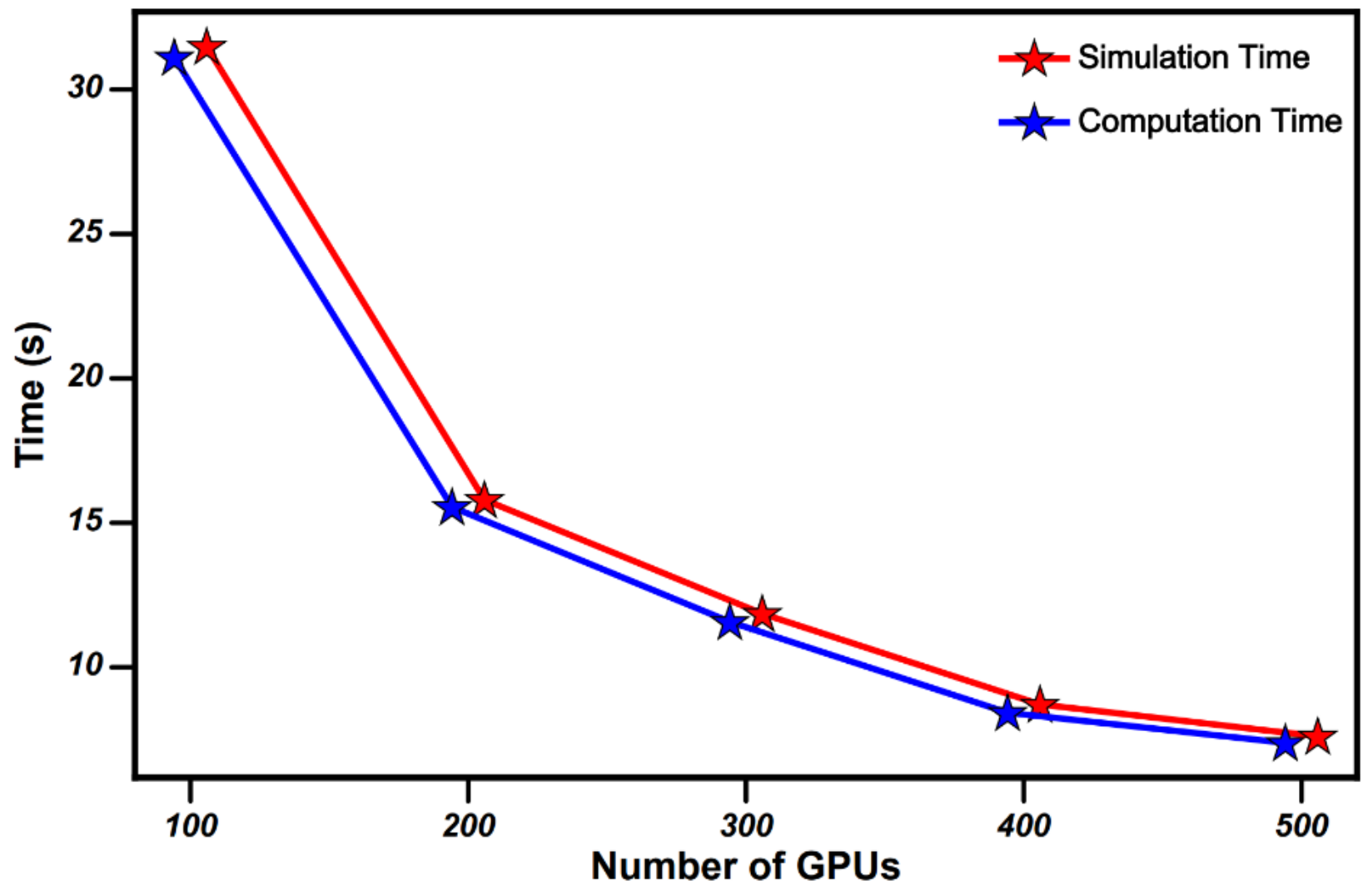

**Supplementary Figure 5. Strong scaling property.** The average time costs of simulation and computation in simulating one biological second of the DB are plotted concerning the number of GPUs: 100, 200, 300, 400 and 500, for networks with 1.5 billion neurons. and varied the number of GPUs: 100, 200, 300, 400, and 500.

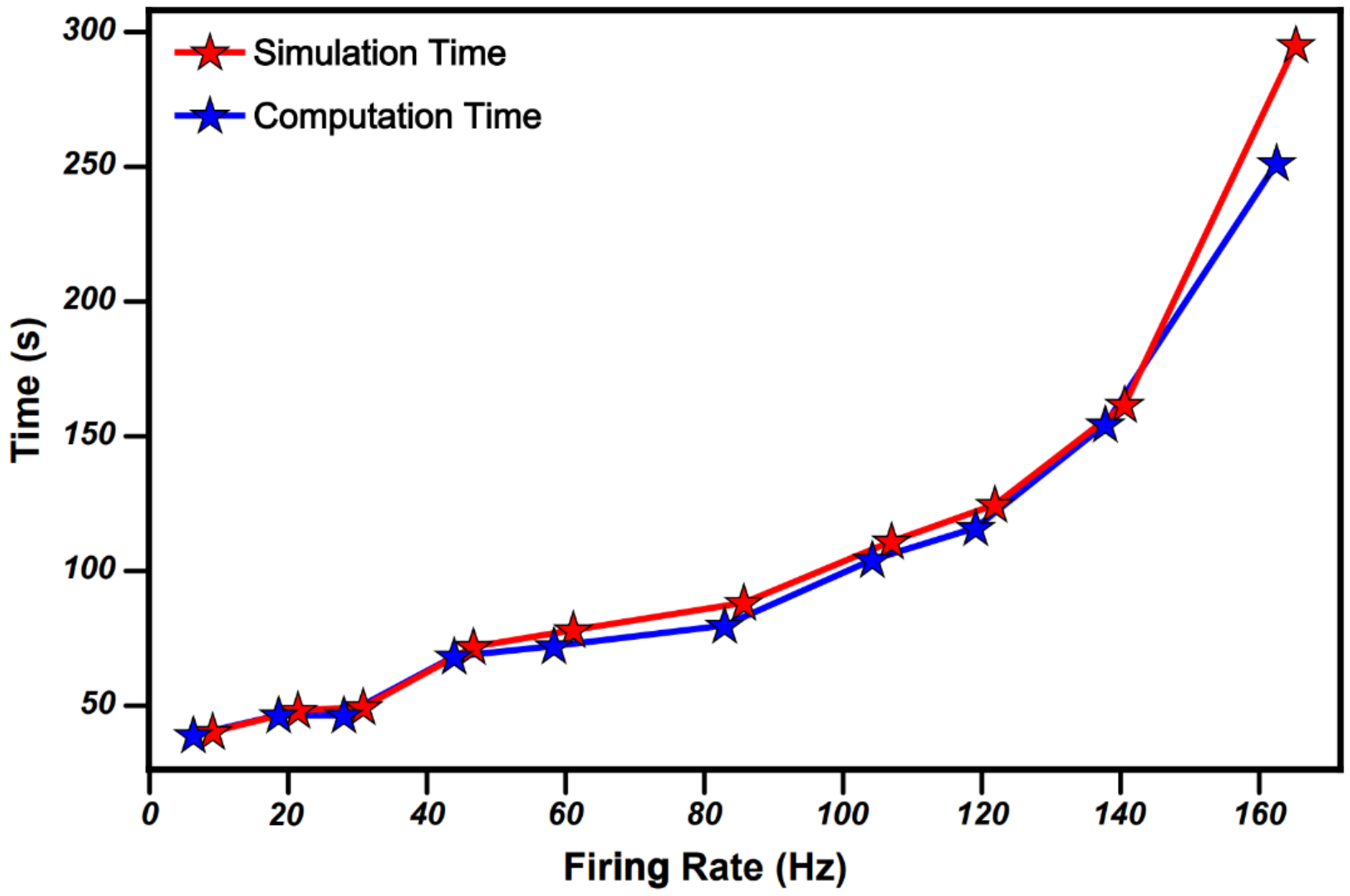

**Supplementary Figure 6. Different firing rates.** The average time costs of simulation and computation in simulating one biological second of the DB are plotted with respect to the firing

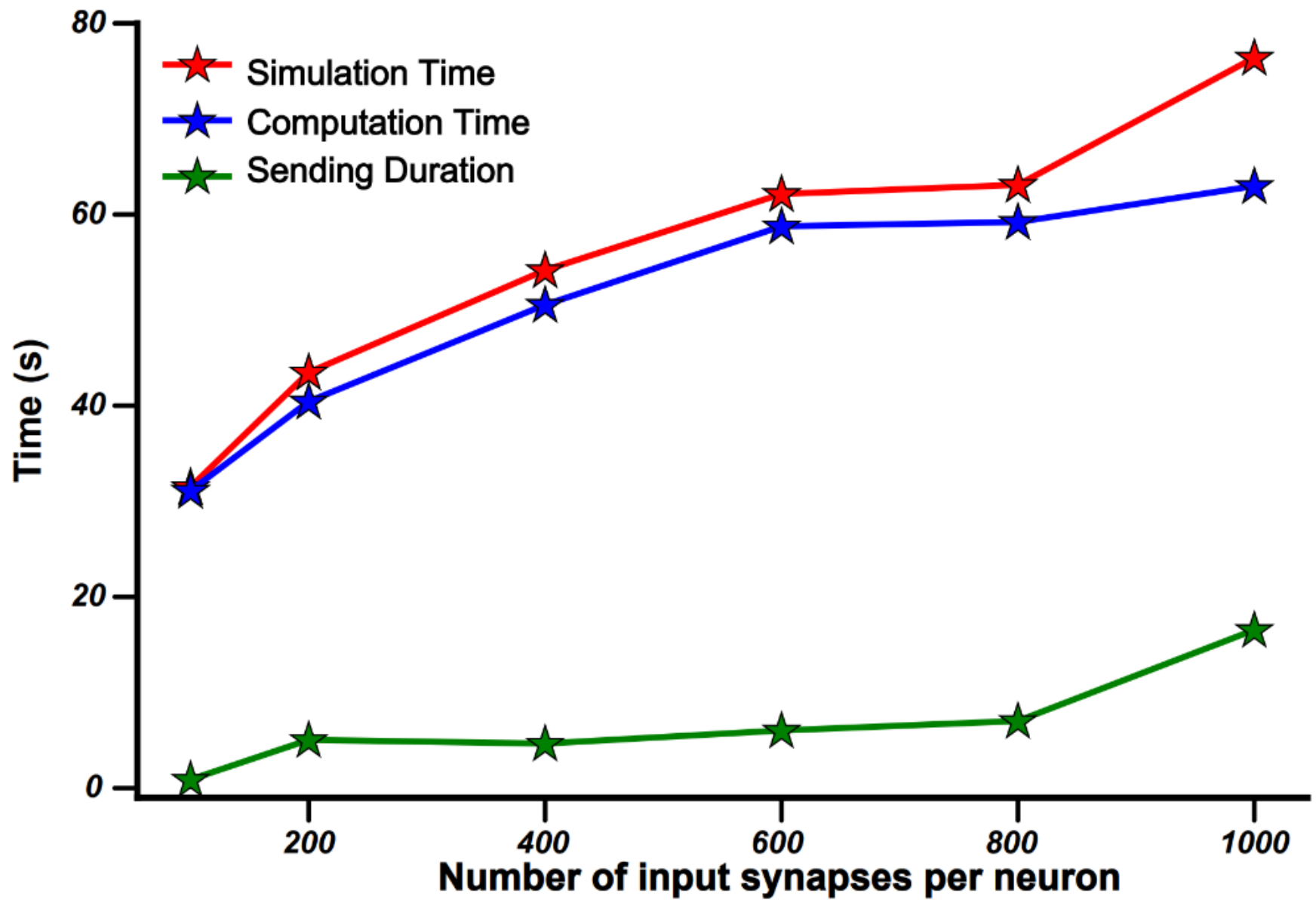

**Supplementary Figure 7. Different synaptic in-degree.** The average time costs of simulation, computation, and sending in simulating one biological second of the DB are plotted concerning the in-degrees of the network and the number of GPUs.

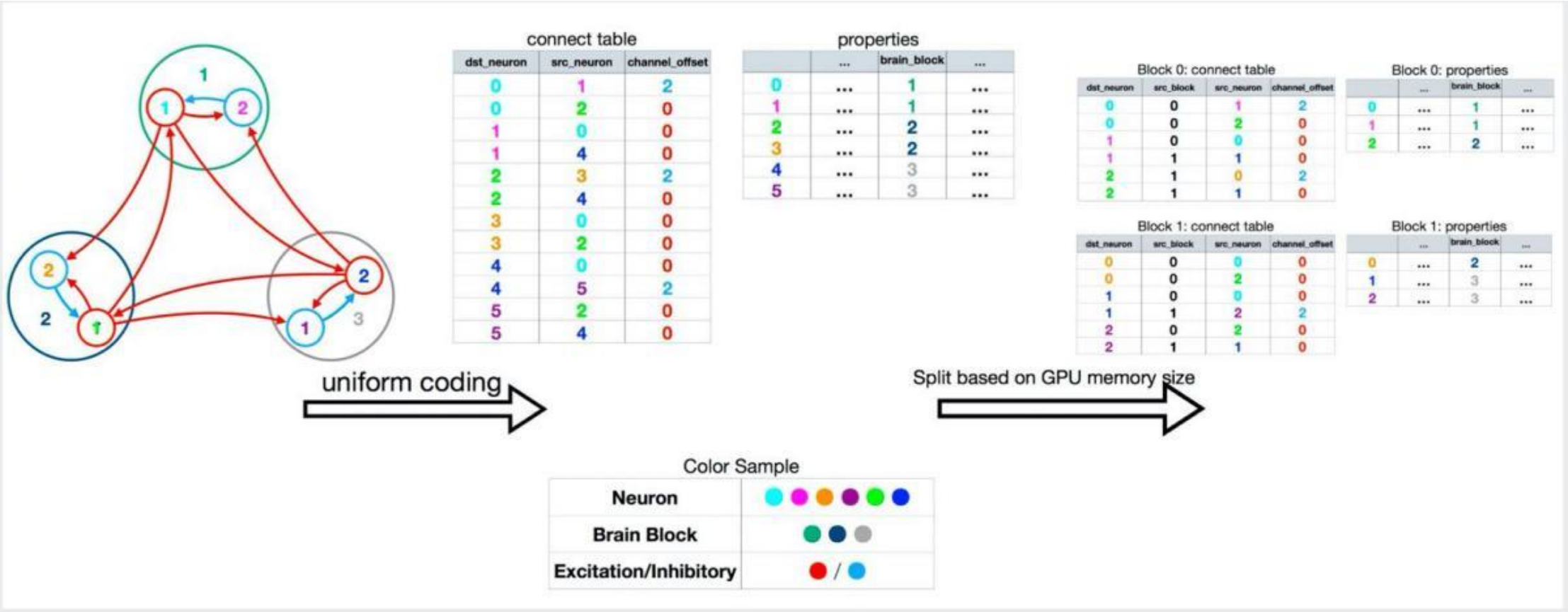

**Supplementary Figure 8. The link table and spike communication organization.**

| **Supplementary Table 1：Statistical indicator and notations** | | | | | | |
|---|---|---|---|---|---|---|
| Statistical indicator | Computation before communication | Computation after communication | Computing ending | Sending copy starting | Sending copy ending | Receiving copy starting |
| Notation | $t_1$ | $t_2$ | $t_3$ | $t_4$ | $t_5$ | $t_6$ |
| Statistical indicator | Receiving copy ending | Sending starting | Sending ending | Receiving starting | Receiving ending | Sending duration inter nodes |
| Notation | $t_7$ | $t_8$ | $t_9$ | $t_{10}$ | $t_{11}$ | $T_{send,inter}$ |
| Statistical indicator | Sending duration intra node | Receiving duration inter nodes | Receiving duration intra node | | | |
| Notation | $T_{send,intra}$ | $T_{rec,inter}$ | $T_{rec,intra}$ | | | |

| Supplementary Table 2: Survey of Large-scale Brain Simulation | | | | | | | |
|---|---|---|---|---|---|---|---|
| Brain simulation | Neuron Nos | Synapse Nos | Mean Firing rates | Real-time factor | Simulation time-step | Connection topology | Model |
| Frye[1] | 8 million | 50.4 billion | 1 Hz | 10 | 1 msec | completely random | Izhikevich model |
| Ananthanarayanan[2] | 16 million | 128 billion | 4.95 Hz | 33.6 | 1 msec | completely random | Izhikevich model |
| Ananthanarayanan[3] | 57.67 million | 461 billion | 7.2 Hz | 65 | 1 msec | completely random | Izhikevich model |
| Ananthanarayanan[4] | 1.617 billion | 8.87 trillion | 19.1 Hz | 643 per Hz | 0.1 msec | anatomical data of a cat's brain | Izhikevich model |
| Helias[5] | 20 million | 225 billion | 6.9 Hz | < 240 | 0.1 msec | - | LIF model |
| Helias[5] | 64 million | 720 billion | 6.9 Hz | < 600 | 0.1 msec | - | LIF model |
| Kunkel[6] | 1.86 billion | 11.1 trillion | - | 2481 | 0.1 msec | - | - |
| Igarashi[7] | 6.04 billion | 24.5 trillion | 4 Hz | 322 | 0.1 msec | neuron distances | LIF model |
| Gosui[8] | 1.048 million | - | 1 Hz | 0.88 | 1 msec | anatomical data of a cat's brain | - |
| Yamazaki[9] | 1 billion | - | - | 0.783 | 1 msec | ditto | - |
| Yamaura[10] | 68 billion | - | - | 416 | 0.1 msec | anatomical data of the human's cerebellar | - |
| Preissl[11] | 65 billion | 16 trillion | 8.1 Hz | 388 | 1 msec | anatomical data of the macaque's brain | LIF model |
| Djurfeldt, Lundqvist et al.[12] | 22 million | 11 billion | 6 Hz | 5942 | - | - | Multicompartmental model neurons |
| Knight[32] | 4.13 million | 24.2 billion | 8Hz (estimated from violin plots) | 504 | 0.1 msec | anatomical data of the macaque's brain | LIF model |
| Wang[33] | 25.6 million | 25.6 billion | 8Hz | 12.175 | 1 msec | completely random | Izhikevich model |
| Our work | 86 billion | 47.8 trillion | 7Hz<br>15Hz<br>30Hz | 65<br>78.8<br>118.8 | 1 msec | sMRI and DTI data, PET synapse data of the human brain combined | LIF model |
| | | | | | | with anatomical cortical microcolumn data of cats' brain | |

| Supplementary Table 3: Brain Label Information for AAL Atlas | | |
|---|---|---|
| Index | Region name | Abbreviation |
| 1,2 | Left/Right Precentral gyrus | PreCG |
| 3,4 | Left/Right Superior frontal gyrus, dorsolateral | SFGdor |
| 5,6 | Left/Right Superior frontal gyrus, orbital part | ORBsup |
| 7,8 | Left/Right Middle frontal gyrus | MFG |
| 9,10 | Left/Right Middle frontal gyrus, orbital part | ORBmid |
| 11,12 | Left/Right Inferior frontal gyrus, opercular part | IFGoperc |
| 13,14 | Left/Right Inferior frontal gyrus, triangular part | IFGtriang |
| 15,16 | Left/Right Inferior frontal gyrus, orbital part | ORBinf |
| 17,18 | Left/Right Rolandic operculum | ROL |
| 19,20 | Left/Right Supplementary motor area | SMA |
| 21,22 | Left/Right Olfactory cortex | OLF |
| 23,24 | Left/Right Superior frontal gyrus, medial | SFGmed |
| 25,26 | Left/Right Superior frontal gyrus, medial orbital | ORBsupmed |
| 27,28 | Left/Right Gyrus rectus | REC |
| 29,30 | Left/Right Insula | INS |
| 31,32 | Left/Right Anterior cingulate and paracingulate gyri | ACG |
| 33,34 | Left/Right Median cingulate and paracingulate gyri | DCG |
| 35,36 | Left/Right Posterior cingulate gyrus | PCG |
| 37,38 | Left/Right Hippocampus | HIP |
| 39,40 | Left/Right Parahippocampal gyrus | PHG |
| 41,42 | Left/Right Amygdala | AMYG |
| 43,44 | Left/Right Calcarine fissure and surrounding cortex | CAL |
| 45,46 | Left/Right Cuneus | CUN |
| 47,48 | Left/Right Lingual gyrus | LING |
| 49,50 | Left/Right Superior occipital gyrus | SOG |
| 51,52 | Left/Right Middle occipital gyrus | MOG |
| 53,54 | Left/Right Inferior occipital gyrus | IOG |
| 55,56 | Left/Right Fusiform gyrus | FFG |
| 57,58 | Left/Right Postcentral gyrus | PoCG |
| 59,60 | Left/Right Superior parietal gyrus | SPG |
| 61,62 | Left/Right Inferior parietal, but supramarginal and angular gyri | IPL |
| 63,64 | Left/Right Supramarginal gyrus | SMG |
| 65,66 | Left/Right Angular gyrus | ANG |
| 67,68 | Left/Right Precuneus | PCUN |
| 69,70 | Left/Right Paracentral lobule | PCL |
| 71,72 | Left/Right Caudate nucleus | CAU |
| 73,74 | Left/Right Lenticular nucleus, putamen | PUT |
| 75,76 | Left/Right Lenticular nucleus, pallidum | PAL |
| 77,78 | Left/Right Thalamus | THA |
| 79,80 | Left/Right Heschl gyrus | HES |
| 81,82 | Left/Right Superior temporal gyrus | STG |
| 83,84 | Left/Right Temporal pole: superior temporal gyrus | TPOsup |
| 85,86 | Left/Right Middle temporal gyrus | MTG |
| 87,88 | Left/Right Temporal pole: middle temporal gyrus | TPOmid |
| 89,90 | Left/Right Inferior temporal gyrus | ITG |

| **Supplementary Table 4: Real-time factor comparison of SNN simulation between NEST and our Digital Brain** | | | | | | |
|---|---|---|---|---|---|---|
| | Neuron Nos. / Mean Firing rate | 2M | 3M | 4M | 5M | 10M |
| NEST | 2.7 HZ | 43.41 | 66.59 | 88.43 | 109.64 | _ |
| | 9.4 HZ | 61.63 | 95.26 | 128.85 | 185.41 | _ |
| Digital Brain (single) | 4.1 HZ | 4.49 | 6.02 | 7.49 | 9.18 | 17.79 |
| | 12.1 HZ | 7.99 | 6.28 | 7.82 | 17.05 | 24.32 |
| Digital Brain (double) | 4.1 HZ | 5.18 | 7.09 | 8.59 | 10.62 | 21.63 |
| | 12.1 HZ | 8.67 | 7.11 | 9.59 | 15.66 | 21.41 |

**Supplementary Table 5: The two tables represent the property and connection for each GPU card.**

Property Table (N * 20)

| Item | Dtype | Interpretation |
|---|---|---|
| $I_{ext}$ | Float32 | External current |
| $P_i$ | Int32 | Population index |
| $C$ | Float32 | Capacitance |
| $T_{ref}$ | Float32 | Refractory period |
| $g_L$ | Float32 | Leaky conductance |
| $V_L$ | Float32 | Leaky Potential |
| $V_{th}$ | Float32 | Threshold potential |
| $V_{reset}$ | Float32 | Reset potential |
| $g_u$ | Float32 | Synaptic conductance (4 channels) |
| $V_u$ | Float32 | Corresponding reversal potential |
| $\tau_u$ | Float32 | Corresponding decay constant |

Connection Table (E * 5)

| Item | Dtype | Interpretation |
|---|---|---|
| The related neuron index of the source | Uint32 | The index of pre-synaptic neurons in this GPU card |
| The related neuron index of the destination | Uint32 | The index of post-synaptic neurons in this GPU card |
| The related card index of the source belongs to | Int16 | The index of the GPU card in which the pre-synaptic neuron located |
| Corresponding weight (2 channels) | Float32 | Weight of AMPA and NMDA for excitatory connection, etc.<br><br>Weight of AMPA and NMDA for excitatory connection, and etc. |

**Supplementary Algorithm 1: Algorithm for parameter initialization**

Input:
Default fire rate: $r_{normal}$;
Maximum fire rate: $r_{max}$;
Channel noise rate: $r_{input}$;
Iteration: $N$;
Output:
Conductance coefficients: $g_u$;

1. Calculate appropriate PSPs: $I_u$;
2. Initialize the coefficient: $g_u^0$;
3. Set $i = 1$
4. **while** $i < N$ or $\bar{I}_u \approx I_u$ **do**
5. run the off-line model
6. obtain the temporary synaptic current: $\bar{I}_u$
7. modify $g_u^i = g_u^{i-1} \times I_u / \bar{I}_u$
8. end while

**Supplementary Algorithm 2: Algorithm for push-based spike integrals**

Input:

All parameters of neurons and graph structure;

Output:

Membrane potentials and spike trains of all neurons;

1. For each neuron, update neuronal state $A_j(t)$;
2. search spiking neurons, and stack them as *SpikeList*;
3. **for** $i \in SpikeList$ **do**
4. update upcoming synaptic inputs to post-synaptic neurons;
5. determine firing neurons and reset to $V_r$ within a refractory time.
6. end for

**Supplementary Algorithm 3: Algorithm for diffusion ensemble Kalman filter**

Input:

BOLD signals: $y_t$, $t = 1...T$;

Number of total samples: $N$;

Total time length: $T$;

Initial state and initial estimation;

Fusion coefficient: $\gamma$;

Dimension of observation state: $L$ (equals to the number of voxels);

Output:

Estimated BOLD signals;

Hidden states and parameters;

1. Draw $N$ samples $x_0^n$ from initial distribution;

2. **for** $t = 1:T$ **do**

3. evolve state $\hat{x}_{t-1}^n$ to $\hat{x}_t^n$ respectively;

4. calculate $\mu_t = \frac{1}{N}\sum \hat{x}_t^n$, $C_t = \frac{1}{N-1}\sum (\hat{x}_t^n - \mu_t)(\hat{x}_t^n - \mu_t)^T$;

5. derive Kalman gain matrix $S_t^l = H_l C_t H_l^T + \Gamma_{o,l}$, $K_t^l = C_t H_l^T (S_t^l)^{-1}$;

6. filter by $x_{t+12}^{n,l} = \hat{x}_t^n + K_t^l (y_t - \varepsilon_o^{n,l} - H_l \hat{x}_t^n)$ for $l = 1...L$;

7. correct $x_t^{n,l}$ based on $x_{t+1/2}^{n,l} : x_t^n = diag(x_{t+1/2}^n \Upsilon(\gamma))$, where $x_{t+1/2}^n = \left[ x_{t+1/2}^{n,l} \right]_l$ and $\left[\Upsilon(\gamma)\right]_{i,j} = \gamma$ if $i = j$ else $\frac{1-\gamma}{L-1}$.

8. end for